\documentclass[twocolumn]{aastex701}

\usepackage{amsmath}
\usepackage{subcaption} 
\begin{document}

\title{Using Ringed Disks to Determine Fundamental Parameters of Planet Formation}

\author[0009-0005-1880-4052,sname='Wu',gname='Olivia']{Olivia Wu}
\affiliation{Department of Mathematics, University of California, San Diego, La Jolla, CA 92093-0112, USA}
\affiliation{Department of Astronomy \& Astrophysics, University of California, San Diego, La Jolla, CA 92093-0424, USA}
\email[show]{odwu@ucsd.edu} 

\author[0009-0008-8711-5367]{Luca Camarra}
\affiliation{PhysicsGraph, Montr\'eal, QC, Canada}
\affiliation{Department of Physics, McGill University, 3600 rue University, Montr\'eal, QC H3A 2T8, Canada}
\email[show]{camarraluca@gmail.com}

\author[orcid=0000-0002-1228-9820,gname=Eve,sname=Lee]{Eve J. Lee} 
\affiliation{Department of Astronomy \& Astrophysics, University of California, San Diego, La Jolla, CA 92093-0424, USA}
\affiliation{Department of Physics, McGill University, 3600 rue University, Montr\'eal, QC H3A 2T8, Canada}
\email[show]{evelee@ucsd.edu}

\begin{abstract}

Two non-dimensional parameters, Stokes number (St) and turbulent $\alpha$, critically control the early stage of planet formation. In spite of their importance, these two numbers are notoriously difficult to obtain and remain largely unconstrained, except in ringed disks. In particular, \citet{Lee2024} showed that ringed disks can be uniquely leveraged to derive local St and $\alpha$ through a simple model combining dust radial equation of motion with the measured distribution of dust masses in Class 0/I disks. We apply their model to the currently known census of ringed disks, finding 19 viable rings in 10 systems, which more than doubles the previous sample. Similar to previous findings, we obtain generally low St$\sim 10^{-4}$--$10^{-2}$ and $\alpha \sim 10^{-5}$--10$^{-2}$ across a wide range of system parameters, consistent with all the disks that we study to be rich in gas and the dust-to-gas ratio of the rings to be too low to generate planetesimals. Disks around low mass host stars likely have leakier traps, broadly consistent with massive, gap-carving giants to be rare around low mass hosts, if the rings are generated by planets.
\end{abstract}

\section{Introduction} \label{sec:Introduction}

The early phase of planet formation is sensitively determined by the properties of the solid and gas that make up the protoplanetary disk. The initial growth of planetary cores is expected to proceed by pebble accretion \citep[e.g.,][]{Ormel10,Lambrechts12,Ormel17} whereby particles of Stokes number St $< 1$ settle onto an accreting body via aerodynamic drag. Once triggered, the rate of growth by pebble accretion is determined in part by the value of St, with smaller St slowing down the growth.

Another factor determining the solid growth is the Shakura-Sunyaev turbulent parameter $\alpha$. Large $\alpha$ (corresponding to stronger diffusive turbulence) can puff up the dust disk and the resulting low local solid density slows down pebble accretion. Following pebble accretion under an evolving background disk of solids undergoing a radial drift \citep[e.g.,][]{Weidenschilling77}, \citet{Lin18} report that a necessary (but not sufficient) condition for the rapid coagulation of massive cores to nucleate a gas giant is St/$\alpha >$ 1.

The $\alpha$ parameter also affects how much the planets may migrate after they form. Both analytic and numerical investigations find that once $\alpha \lesssim 10^{-4}$--$10^{-3}$, disk-induced Type I migration can stall prematurely as the perturbed gas interior to the planet's orbit piles up before it can be diffused. The resulting increase in the inner Lindblad torque can drive the net torque on the planet to zero \citep[e.g.,][]{Rafikov02,Li09,Yu10,Fung17,Fung18}. 

In sum, the two non-dimensional parameters St and $\alpha$ critically determine both the mass growth of rocky cores and the radial emplacement of the planet within the protoplanetary disks, implying that the fundamental observables of exoplanets---masses, radii, and orbital periods---are all governed by St and $\alpha$.

Direct measurements of St and $\alpha$, however, are challenging. Under the Epstein drag, Stokes number is defined as
\begin{equation}
    {\rm St} = \Omega\frac{\rho_s s}{\rho_g c_s}
    \label{eq:St}
\end{equation}
where $\Omega \equiv \sqrt{GM_\star /a^3}$ is the Keplerian orbital frequency, $G$ is the gravitational constant, $M_\star$ is the mass of the host star, $a$ is the orbital distance, $\rho_s$ is the material density of the grain, $s$ is the size of the grain, $\rho_g$ is the volumetric mass density of disk gas, $c_s \equiv \sqrt{kT/\mu m_H}$ is the sound speed, $k$ is the Boltzmann constant, $T$ is the midplane temperature, $\mu=2.3$ is the gas mean molecular weight, and $m_H$ is the Hydrogen atomic mass. Of all the listed parameters here, the density $\rho_g$ is the most problematic and least constrained as the gas measurements are limited, and even when CO measurements are available \citep[e.g.,][]{Zhang21}, conversion to the total gas mass depends on the uncertain CO-H$_2$ factor.

The subsonic nature of turbulence in protoplanetary disks renders the measurement of $\alpha$ equally challenging. Where reports are made---based on the depth of the gap in dust emission \citep{Pinte2016} or the non-thermal broadening of molecular lines \citep{Flaherty15,Flaherty17}---$\alpha \lesssim 10^{-3}$ over a wide range of vertical depth and radial extent. 

Alternatively, the dust ring substructures in protoplanetary disks seen in Atacama Large Millimeter/submillimeter Array (ALMA) have also been used to constrain St and $\alpha$. The rings are commonly found \citep[e.g.,][]{ALMA2015,Andrews18,Bae2023} and their observations are cheaper than high resolution gas kinematics. Modeling the gas pressure perturbation as a Gaussian of width $w$ and considering the dust ring to have reached a steady state equilibrium between the aerodynamic drift that collects the particles at the center of the pressure bump and the turbulent diffusion, the width of the dust ring can be expressed as \citep{Dullemond18}
\begin{equation}
    w_{\rm d} = w\left[1+\left(\frac{{\rm Sc}{\rm St}}{\alpha}\right)\right]^{-1/2}
    \label{eq:dust-ring-width}
\end{equation}
where ${\rm Sc} = 1$ is the Schmidt number taken as 1 for simplicity \citep[see also][]{Johansen05}. Setting $w$ to the local gas disk scale height $H \equiv c_s / \Omega$ and plugging in the measured $w_{\rm d}$ from the resolved dust rings, equation \ref{eq:dust-ring-width} can be inverted to obtain the local ratio St/$\alpha$. Further procedures are required to separately determine St and $\alpha$.

To break the degeneracy between St and $\alpha$, \citet{Rosotti20} set the maximum St to be that determined by fragmentation, reasoning that any larger grains would be ground down by the relative velocity between the grains set by their coupling to gas turbulence. For given St and $\alpha$, maximum St places an upper limit on $\alpha$. The fragmentation velocity however is highly uncertain and varies sensitively on the grain chemical composition and porosity \citep[e.g.,][]{Musiolik19,Kimura20}, both of which are not well known.

\citet{Lee2024} devised a different approach that does not rely on uncertain fragmentation velocities. Instead, they integrated the dust radial equation of motion backward in time to solve for the appropriate combination of St and $\alpha$ that matches the total dust mass flux inside the rings given the distribution of initial dust mass reservoir from Class 0/I disks \citep{Tobin20}. Focusing on five systems with a total of eight well-resolved dust rings from the DSHARP survey \citep{Dullemond18}, \citet{Lee2024} found generally low St $\sim 10^{-4}$--10$^{-2}$ and $\alpha \sim 10^{-5}$--10$^{-3}$. While such values are lower than what is generally adopted in the literature, their derived St was found to be consistent with the inferred gas surface density for HD 163296 where optically thin line measurements were available \citep{Booth19}. 

In this paper, we expand on the sample studied by \citet{Lee2024} to investigate how St and $\alpha$ change across a larger variety of system parameters such as the disk size, host star mass, age, and the external radiation environment. The paper is organized as follows. Section \ref{sec:Method} outlines our method of deriving St and $\alpha$, our sample selection, and how we compute the ring parameters such as their width and mass. Section \ref{sec:Results} presents our results while Section \ref{sec:disc_concl} concludes and discusses implications of our results.

\section{Method} \label{sec:Method}

\subsection{Resolving St and \texorpdfstring{$\alpha$}{alpha}} \label{subsec:St and Alpha}

We begin by highlighting the key points of the methodology of \citet{Lee2024} which we follow to resolve St and $\alpha$ for our ringed disks. We adopt $\Sigma_g \propto a^{-1}$ as the disk gas surface density motivated by \citet{Zhang21} and 
\begin{equation}
    T = \left( \frac{\phi L_{\star}}{8\pi a^2 \sigma_{sb}} \right)^{1/4}
    \label{eqn:Midplane Temperature}
\end{equation}
as the midplane temperature where $\phi = 0.25$ is the grazing angle of incident starlight appropriate at 10s of au \citep{Chiang1997, Dullemond2001},\footnote{We note that our adopted flaring angle $\phi$ is larger than that used by \citet{Dullemond18} by a factor of $\sim$10. Our adoption changes our estimated mass of the ring previously studied by \citet{Dullemond18} by factors of $\sim$2, which we discuss in more detail in Section \ref{subsec:Ring Mass}.} $L_{\star}$ is the luminosity of the central star, and $\sigma_{sb}$ is the Stefan-Boltzmann constant, appropriate for irradiation-heated disk which is a good approximation at wide orbits where the dust rings are observed.

Next, the dust radial equation of motion (including aerodynamic drag and coupling to the gas motion) is integrated back in time to identify the origin site of the dust grains $a_{0}$ that arrive at the location of the observed ring $a_{\rm ring}$:
\begin{equation}
\begin{split}
    \frac{a_{0}(t')}{a_{\rm ring}} &= -\frac{3}{2} \frac{\alpha}{|\gamma|\rm St_r} \\
    &+ \left( 1 + \frac{3}{2} \frac{\alpha}{|\gamma| \rm St_r} \text{Exp} \left[ \frac{c^2_{s,r}}{v_{k,r}} |\gamma| \text{St}_r \frac{(t_{\rm age} - t')}{a_{\rm ring}} \right] \right),
\end{split}
\label{eqn:Dust Radial Motion}
\end{equation}
where $\gamma =-11/4$ is the non-dimensional radial pressure gradient of the disk based on our adopted disk profile, $t_{\rm age}$ is the age of the system, and $v_k = \Omega a$ is the Keplerian velocity. The subscript r represents calculations at $a_{\rm ring}$. The above equation adopts constant grain size so that ${\rm St} \propto a$. 

Our choice of constant grain size is motivated by our usage of ALMA images whose wavelengths are more direct probe of grain size rather than St. However, we implicitly ignore the effect of grain growth and fragmentation whose detailed implementation is beyond the scope of this paper. We note that assuming spatially constant St---which is consistent with accounting for grain growth as they drift in such that the size $\propto a^{-1}$---makes negligible difference in the final St and $\alpha$ \citep[see][their Figure 2]{Lee2024}. We explore the case of a more rapid growth of grains (St $\propto a^{-2}$ in Appendix \ref{sec:app-grain-growth} and find that our computation of St and $\alpha$ remain remarkably stable. We therefore continue with our assumption of constant grain size.

The dust rings are interpreted as the local trap of dust particles that drift from the outer region so we can express its mass as
\begin{equation}
M_{\rm ring}(t) = M_0 \epsilon_{\rm trap} \left[ \left( \frac{a_0(t_{\rm age} - t)}{\text{1 au}} \right)^\beta - \left( \frac{a_{\rm ring}}{\text{1 au}} \right)^\beta \right],
\label{eqn:Mass in Ring at time t}
\end{equation}
where $M_0$ and $\beta$ describe the initial disk mass distribution $M_{\rm disk} = M_0 (a/{\rm 1\, au})^\beta$, and $\epsilon_{\rm trap}$ is the trapping efficiency defined as the ratio of dust mass contained within the ring over the mass of the dust being supplied to the ring. Following \citet{Lee2024}, we use $M_0=35.6M_\oplus$ and $\beta=0.57$ obtained from a least-squares fit to the mass and radii measurements of Class 0/I disks in Orion \citep{Tobin2020}. We vary $\epsilon_{\rm trap} \in [0.3, 0.6, 1]$ given its uncertainty and variation \citep[e.g.,][]{Lee2022a, Huang2025}.

For systems with multiple rings, we account for the effect of dust mass budget for the inner ring(s) being affected by filtering of the outer ring(s). We first check if $a_0$ of each inner ring is larger than $a_{\rm ring}$ of its relative outer ring(s). If this condition is true, we apply the following correction to $M_0$ recursively:
\begin{equation}
M_{0,\rm new} \left( \frac{a_{0,\rm out}}{\text{1 au}}\right)^\beta = M_{0,\rm orig} \left( \frac{a_{0,\rm out}}{\text{1 au}}\right)^\beta - M_{\rm ring,out},
\end{equation}
where $M_{0,\rm orig}$ is the original inner $M_0$ (starting with 35.6$M_\oplus$ at the first iteration), $M_{\rm ring,out}$ is the mass of the outer ring, $a_{0,\rm out}$ is the outer ring's $a_0$, and $M_{0,\rm new}$ is the corrected $M_0$ \citep{Lee2024}. For example, if $a_0$ of the innermost ring of a 3-ringed system surpasses the $a_{\rm ring}$ of the outermost ring, we first calculate the corrected $M_0$ for the middle ring using the outer ring. We then use the corrected middle $M_0$ to resolve the $M_0$ for the inner ring. If the innermost ring's $a_0$ surpasses the middle ring $a_{\rm ring}$ but is within the outermost $a_{\rm ring}$, then the same correction is applied but only between the two inner rings. 

We combine equations \ref{eqn:Dust Radial Motion} and \ref{eqn:Mass in Ring at time t} to directly solve for St:
\begin{align}
    {\rm St}_{\rm r} &= \left(\frac{v_{k,r}}{c_{s,r}^2}\right)\frac{a_{\rm ring}}{t_{\rm age}|\gamma|} \nonumber \\
    &\times\log\left(\frac{\left(\frac{M_{\rm ring}}{M_0\epsilon_{\rm trap}}+\left(\frac{a_{\rm ring}}{{\rm 1 AU}}\right)^\beta\right)^{1/\beta} + \frac{3}{2}\frac{\alpha}{|\gamma|{\rm St}_r}\left(\frac{a_{\rm ring}}{1 \,{\rm AU}}\right)}{\left(\frac{a_{\rm ring}}{\rm 1 AU}\right)+\frac{3}{2}\frac{\alpha}{|\gamma|{\rm St_r}}\left(\frac{a_{\rm ring}}{\rm 1 AU}\right)}\right) \nonumber \\
    \label{eq:St_r}
\end{align}
and $\alpha$ is obtained with equation \ref{eq:dust-ring-width}.

Under our procedure, the values of St and $\alpha$ depend logarithmically on $M_0$, the initial dust mass budget, which vary by approximately an order of magnitude \citep[see][their Figure 1]{Lee2024}. We therefore derive $M_0$ and $\beta$ for each of the disk mass percentiles calculated within each radius bin (see \citealt{Lee2024} for more detail). We then calculate St and $\alpha$ over the range of $M_0$ and $\beta$, scanning incrementally in disk mass percentiles from 0th to 100th in steps of 1. Here, the 0th percentile corresponds to the minimum datapoint for each radius bin whereas the 100th percentile to the maximum datapoint for each radius bin.

There are physical limits to St. First, since St $\propto \Sigma_g^{-1}$ (c.f.~equation \ref{eq:St}), too low a St would imply a disk that may be dense enough to be gravitationally unstable, which places a lower limit on St:
\begin{equation}
{\rm St} > \frac{\pi G \rho_s s}{c_s\Omega},
\label{eq:St_Q}
\end{equation}
where we adopt $\rho_s = \text{1 g cm}^{-3}$ and s$= \lambda/2\pi$ with $\lambda$ being the wavelength of observation. Second, the upper limit on St is set by enforcing $a_0$ to stay within 1000 au. We report these lower and upper limits on St and the corresponding $\alpha$ and disk mass percentiles.\footnote{Our adopted minimum St for each ring may not exactly be that obtained from equation \ref{eq:St_Q} as we choose the closest St numerically obtained for each of the disk mass percentile described above.} 

For systems with multiple rings, after applying these limits for each of the rings, we take the smallest upper bound on $M_0$ (minimum St) and the greatest lower bound on the $M_0$ (maximum St) among the member rings and apply these limits consistently for all the rings in a given system.

\subsection{Sample Selection} \label{subsec:Sample Selection}

To build a sample of well-resolved dust rings, we start with \citep[][PP7 from hereon, their Table 1]{Bae2023}, the Ophiuchus DIsc Survey Employing ALMA (ODISEA) \citep{Cieza2021}, disks around Taurus M stars \citep{Shi2024}, and within $\sigma$ Orionis systems \citep{Huang2024}.

PP7 lists a total of 62 systems. We first remove rings and sources with substructures measured only in infrared as we only consider millimeter sized dust grains in our analysis. This first check removes the system IRS48/WLY2-48. 
Both millimeter and infrared measurements are available for systems LkCa15, PDS70, SR 21, HD163296, and J1604-2130 for which we choose the millimeter measurements. 
We continue to remove sources if the full width half maximum (FWHM) and stellar properties were not listed. 
The rings DoAr25 B-86, Elias24 B-123, HD163296 B-159, TWHya B-52, RULup B-17 and B-50 and the systems DoAr33, ROX27/WSB52, CSCha, V1094Sco, 2MASS J16090141-3925119, T4/SY CHa, SZ Cha, J1604-2130, and Sz118 are removed for not having measurements for FWHM. The system WLY2-63/IRS63 is further removed for not having measurements for its host star's mass and luminosity. 

By visual inspection of their continuum maps, J1680 (2MASS J16083070-3828268) and Sz111 do not appear to have visible ``sharp'' rings so we do not consider them \citep{Villenave2019, Ansdell2018}. We further eliminate HD 142527 as its substructure is a horseshoe and the disk is circumbinary \citep{Boehler2017} which we do not consider for our final sample of rings. After making these cuts, we are left with 48 systems from PP7.

For a substantial number of the remaining systems, we observe that the ring model's 1-$\sigma$ width is listed as the FWHM. We correct this data, adopting 1-$\sigma$ instead of FWHM as the width of the dust ring $w_{\rm d}$ to the following 23 systems: CITau, CIDA1, CIDA9A, DLTau, DNTau, FTTau, GOTau, IPTau, IQTau, LkCa15, MHO6, MWC480/HD31648, RYTau, CQTau, HD163296, HPCha, WWCha, PDS70, Sz91, Sz123A, SAO206462/HD135344B, J1610, and HD100546. 

As we adopt a larger flaring angle than that used by \citet{Dullemond18} which affects $T$, we rederive the ring parameters through our own Gaussian fits to the radial intensity profiles that are publicly available. Deriving our own ring properties also safeguards us from the propagation of rounding errors.
For systems DoAr 25, Elias 20, Elias 24, HD163296, IM Lup, GW Lup \citep{Huang2018}, AS 209 \citep{Guzman2018}, and TW Hya \citep{Andrews2016}, we use deprojected and azimuthally averaged radial intensity profiles and their associated beam dimensions and distances to convert beam intensities to true intensities. We then use the emcee package \citep{emcee} in \texttt{Python} to resolve Gaussian parameters using MCMC fits. We consider two Gaussian models depending on the shape of the ring(s). 

\begin{figure}
    \centering
    \includegraphics[width=7.5cm]{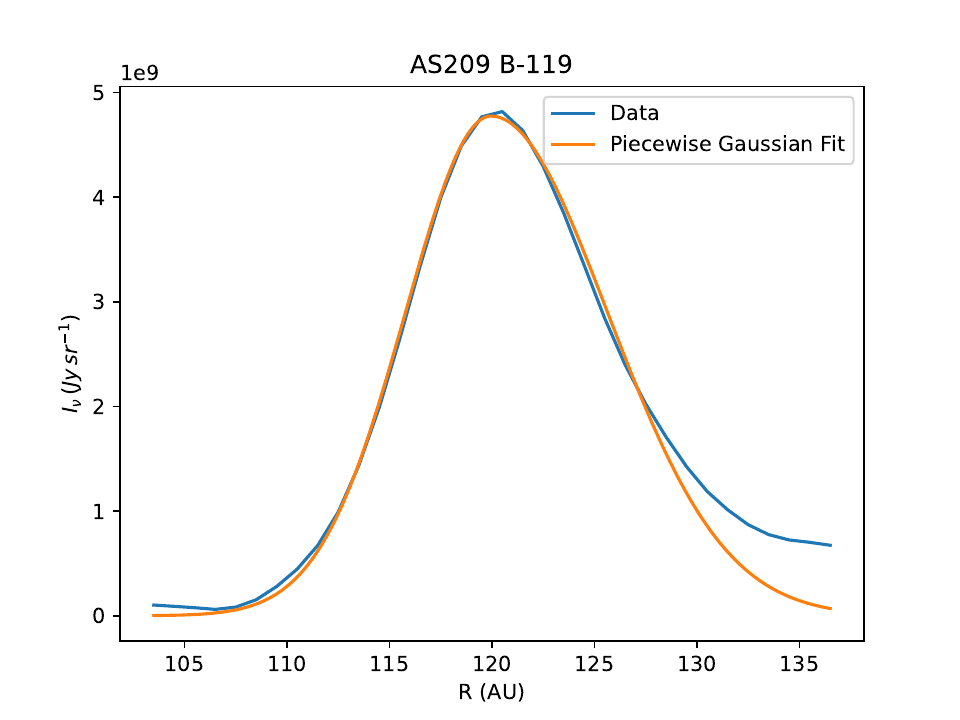}
    \caption{AS209 B-119 as an example for the asymmetric Gaussian fit. The blue line is the deprojected radial intensity data from \citet{Andrews2018a}. The orange line is our best fit to the data, obtaining deconvolved widths $w_{\rm L}=3.64 \text{ AU}$ and $w_{\rm R}=	5.31\text{ AU}$ (equation \ref{eqn:Piecewise Gaussian Fit}).}
    \label{fig:Asymmetric Gaussian Fit Example}
\end{figure}

For a ring that is single or clearly separated from other rings in a given system, we fit an asymmetric Gaussian model to the intensity profile (see Figure \ref{fig:Asymmetric Gaussian Fit Example}):
\begin{equation}
    I_\nu = 
    \begin{cases} 
    \begin{split}
    \text{A exp}{\left[ -\frac{1}{2}\left( \frac{a-a_{\rm ring}}{w_{\rm L}}\right)^2 \right]} & \text{if } a < a_{\rm ring} \\
    \text{A exp}{\left[ -\frac{1}{2}\left( \frac{a-a_{\rm ring}}{w_{\rm R}}\right)^2 \right]} & \text{if } a > a_{\rm ring}
    \end{split}
    \end{cases}
    \label{eqn:Piecewise Gaussian Fit}
,
\end{equation}
where $w_{\text{L}}$ and $w_{\text{R}}$ are the left and right Gaussian widths, and A is the intensity amplitude. Between the two widths, we favor the larger one to make a conservative selection on the sharpness of the ring.

\begin{figure}
    \centering
    \includegraphics[width=7.5cm]{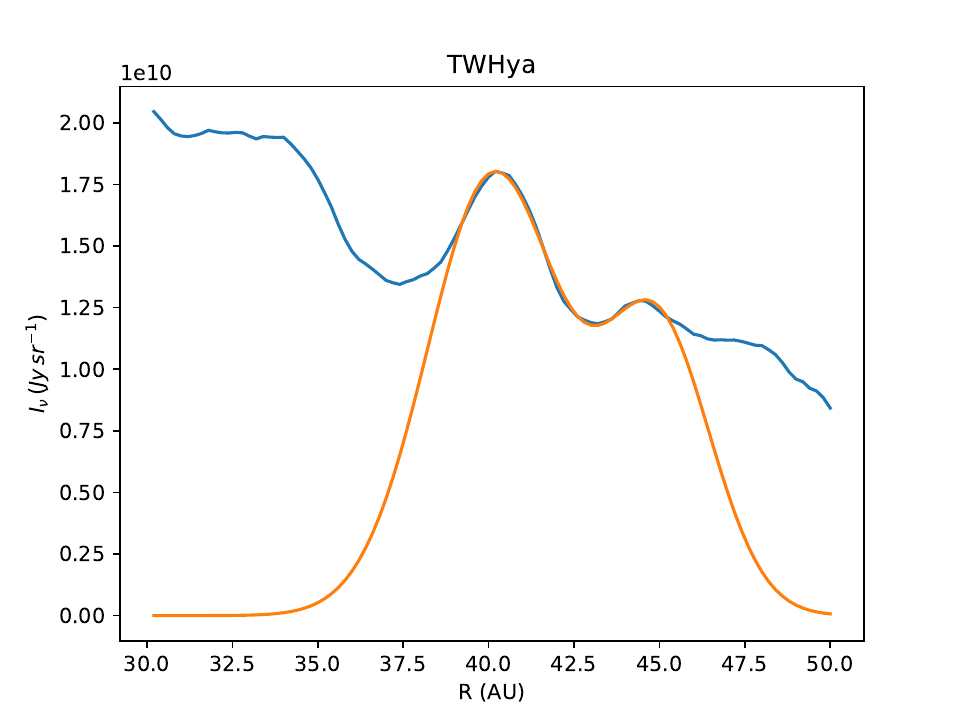}
    \caption{TW Hya as an example for the double Gaussian model fit. The blue line is the deprojected radial intensity data from \citet{Andrews2016} showing B-40 and B-45. The orange line represents our best fit, obtaining de-convolved widths $w_1=1.86 \text{ AU}$ and $w_2=1.5 \text{ AU}$ (equation \ref{eqn:Double Gaussian Fit}).}
    \label{fig:Double Gaussian Fit Example}
\end{figure}

For TW Hya, DoAr 25, Elias 20, and AS 209, we observe double peaks in the radial intensity profiles, suggesting two rings are nearby. We therefore fit the following double Gaussian model (see Figure \ref{fig:Double Gaussian Fit Example}): 
\begin{equation}
\begin{split}
        I_\nu = \text{ A}_1\text{ exp}{\left[ -\frac{1}{2}\left( \frac{a-a_{\rm ring,1}}{w_{1}}\right)^2 \right]} \\ 
        + \text{ A}_2\text{ exp}{\left[ -\frac{1}{2}\left( \frac{a-a_{\rm ring,2}}{w_{2}}\right)^2 \right]},
\end{split}
\label{eqn:Double Gaussian Fit}
\end{equation}
where ($\text{A}_1$, $\text{A}_2$), ($w_{1}$, $w_{2}$), and ($\rm a_{\rm ring,1}$, $\rm a_{\rm ring,2}$) are the amplitudes, widths, and orbital distances for the first and second ring respectively. From this fitting, we eliminate the AS 209 rings B-97 and B-141 because we do not resolve reasonable fits to the radial intensity profiles and visually do not observe rings in their radio continuum maps \citep{Huang2018}.

The intensity models we fit are the product of a convolution of the synthesized ALMA beams and the underlying intensity profiles. Assuming a Gaussian beam and a Gaussian intensity profile in the vicinity of the ring, we obtain the underlying width of the dust ring using the rule of the convolution of two Gaussians,
\begin{equation}
    w_\mathrm{d}=\sqrt{\sigma^2-\sigma_b^2}
\end{equation}
where $w_\mathrm{d}$ is the de-convolved width of the dust ring, $\sigma$ is the width of the ring obtained from the Gaussian fits, and $\sigma_b$ is the width of the beam as a standard deviation in astronomical units \citep{Dullemond2018}. We also correct for the distortion of the beam width due to the ellipticity of the beam and inclination of the disk as described in Appendix H of \citet{Dullemond2018}:
\begin{equation}
    \sigma_b=\sqrt{\frac{\sigma_\mathrm{min}\sigma_\mathrm{maj}}{ | \cos i |}}
\end{equation}
where $\sigma_\mathrm{min}$ and $\sigma_\mathrm{maj}$ are the minor and major axes of the elliptical beam, and $i$ is the inclination of the disk. This approximation holds in most cases, except when the disk is highly inclined and the beam is strongly elliptic. 

Next, PP7 listed stellar parameters for a number of systems that either do not agree with what is quoted or are not available in the corresponding references. We correct these parameters to that listed in the literature. Additionally, we add data for system age when available.
For SZ 123A, we take stellar properties, distance, and age from \citet{Guerra-Alvarado2025}. For WW Cha, TW Hya, SZ 91, and SAO206462/HD 135344B, we correct stellar mass, stellar luminosity, distance and age with data quoted in \citet{Andrews2018a}. For AATau, we take the system age from \citet{Gudel2007} and for SR 21, age is taken from \citet{Andrews2018a}. Distance for V1247 Ori is taken from \citet{Ren2024} and distance for SR 21 from \citet{Andrews2018a}.

We additionally consider systems in $\sigma$ Ori which are expected to be in harsh UV environment \citep{Huang2024}. Exclusive to the $\sigma$ Ori disks, we make corrections to the midplane temperature accounting for the external radiation these systems experience. We compute the external flux from nearby bright stars as $\rm F_{\rm ext} \sim 1150 \text{ L}_\odot \text{ pc}^{-2}$ using the luminosity of the nearby bright star $\sigma$ Ori Aa 41,700 $\rm L_\odot$ \citep{Simon-Diaz15} and with projected separations of 1.7 pc \citep{Huang2024}. The new midplane temperature is computed using
\begin{equation}
    \sigma_{\rm sb}T^4 = F_\star + F_{\rm ext}
    \label{eq:Tmid_sigori}
\end{equation}
where $F_\star = \phi L_\star / 8\pi a^2$ from equation \ref{eqn:Midplane Temperature}. Comparing our updated temperature to the temperature calculated purely from irradiation from the central star, we see percentage differences ranging from only 0.13\% to 1.4\%, confirming the statement of \citet{Huang2024} that the effect of external {\it bolometric} flux on $\sigma$ Ori disks is minor.

\subsection{List of Sharp Dust Rings}

In our analysis, we search for ``sharp'' rings defined as where $w_{\rm d} < H/\sqrt{2}$, equivalent to St/$\alpha > 1$ in equation \ref{eq:dust-ring-width} whose heuristic interpretation is that the sharp rings arise from the balance between the aerodynamic drag and the turbulent diffusion.
For the ODISEA dataset, the uncertainty in the location and the width of the rings was substantial so we apply the sharp ring criterion for both its optimistic (large $a_{\rm ring}$ and small $\sigma$) and pessimistic (small $a_{\rm ring}$ and large $\sigma$) scenarios. 
Using our criterion, we resolve 14 rings from PP7, no rings from ODISEA, 2M0412 B-114, and SO 1274 B-50, B-66, B-88, and B-119 as sharp.
Under this definition of ``sharp'' rings, we note that DSHARP rings HD 163296 B-67, HD 143006 B-41, and HD 143006 B-65 do not qualify, so we exclude them from our final dataset. We include rings AS 209 B-74, AS 209 B-120, Elias 24 B-77, HD 163296 B-100, and GW Lup B-85 as they all satisfy the criterion.
Our final list includes 19 rings in 10 systems. Table \ref{tab:Ring List} lists these systems, their corresponding rings, stellar parameters, and dust ring parameters. 

\begin{deluxetable*}{lcCCcCCCCCC}
\tablewidth{0pt}
\tablecaption{Selection of Well-resolved Dust Rings Gathered from Literature Search \label{tab:Ring List}}
\tablehead{
\colhead{Source} & \colhead{Name} & \colhead{Wavelength} & \colhead{Age} & \colhead{Reference} & \colhead{$M_{\star}$} & \colhead{$L_{\star}$} & \colhead{$a_{\rm ring}$} & \colhead{$w_d$}  & \colhead{$M_{\text{ring}}$} & \colhead{$\rm St/\alpha_{\rm min}$}\\
\colhead{} & \colhead{} & \colhead{(mm)} & \colhead{(Myr)} & \colhead{} & \colhead{$(M_{\odot})$} & \colhead{$(L_{\odot})$} & \colhead{(au)} & \colhead{(au)} & \colhead{$(M_{\oplus})$} 
}
\decimalcolnumbers
\startdata
GO Tau & B-74 & 2.90 & 2.20 & 1, 3 & 0.36 & 0.21 & 73.8 & 1.99 & 13.3 & 20.0 \\
GO Tau & B-111 & 2.90 & 2.20 & 1, 3 & 0.36 & 0.21 & 109.0 & 9.40 & 16.7 & 1.53 \\
HL Tau & B-49 & 1.30 & 1.0 & 2, 4, 5, 10 & 1.7 & 11.0 & 49.0 & 1.78 & 30.0 & 8.11 \\
HL Tau & B-72 & 1.30 & 1.0 & 2, 4, 5, 10 & 1.7 & 11.0 & 72.2 & 1.74 & 7.0 & 24.2 \\
HL Tau & B-85 & 1.30 & 1.0 & 2, 4, 5, 10 & 1.7 & 11.0 & 85.4 & 5.10 & 101.0 & 3.48 \\
DoAr 25 & B-111 & 1.25 & 2.0 & 2, 4 & 0.95 & 0.95 & 111.0 & 6.73 & 40.88 & 1.84 \\
DoAr 25 & B-137 & 1.25 & 2.0 & 2, 4 & 0.95 & 0.95 & 138.0 & 9.91 & 51.19 & 1.25 \\
Elias 24 & B-77 & 1.25 & 0.20 & 2, 4 & 0.78 & 6.02 & 76.5 & 4.47 & 15.64 & 3.86 \\
AS 209 & B-74 & 1.25 & 1.0 & 2, 4 & 0.83 & 1.41 & 74.2 & 3.62 & 12.11 & 3.50 \\
AS 209 & B-120 & 1.25 & 1.0 & 2, 4 & 0.83 & 1.41 & 120.0 & 5.15 & 25.49 & 6.38 \\
HD 163296 & B-100 & 1.25 & 7.94 & 2, 4 & 2.0 & 17.0 & 99.6 & 5.55 & 19.03 & 2.10 \\
TW Hya & B-40 & 0.87 & 6.30 & 4, 6 & 0.81 & 0.34 & 40.2 & 1.79 & 3.76 & 1.86 \\
TW Hya & B-45 & 0.87 & 6.30 & 4, 6 & 0.81 & 0.34 & 44.9 & 1.45 & 2.37 & 4.71 \\
GW Lup & B-86 & 1.25 & 2.0 & 1, 4 & 0.46 & 0.33 & 85.5 & 5.93 & 14.20 & 2.0 \\\
2M0412 & B-114 & 1.30 & 4.50 & 7, 8 & 0.30 & 0.13 & 114.0 & 9.38 & 15.26 & 1.96 \\
SO 1274 & B-50 & 1.30 & 2.0 & 9 & 0.64 & 0.68 & 49.7 & 2.05 & 4.34 & 4.57 \\
SO 1274 & B-66 & 1.30 & 2.0 & 9 & 0.64 & 0.68 & 65.6 & 2.55 & 6.40 & 6.25 \\
SO 1274 & B-88 & 1.30 & 2.0 & 9 & 0.64 & 0.68 & 88.2 & 3.05 & 7.78 & 9.62 \\
SO 1274 & B-119 & 1.30 & 2.0 & 9 & 0.64 & 0.68 & 119.0 & 2.25 & 11.7 & 41.0 \\
\enddata
\tablecomments{Column(1): name of the source. Column(2): name of the ring. Column(3): wavelength dust ring was observed at by ALMA. Column(4): age of the system. Column(5): literature reference. Column(6): mass of the central star. Column(7): luminosity of central star. Column(8): orbital distance. Column(9): width of dust ring. Column(10): measured dust mass of the ring. Column(11): minimum St vs $\alpha$ ratio. 
References: (1) \citet{Andrews2018a}, (2) \citet{Andrews2018b}, (3) \citet{Long2019}, (4) \citet{Huang2018}, (5) \citet{ALMAPartnership2015}, (6) \citet{Sokal2018}, (7) \citet{Long2023}, (8) \citet{Shi2024}, (9) \citet{Huang2024}, (10) \citet{Pinte2016}.}
\end{deluxetable*}

\begin{deluxetable*}{cccccccccccc}
\tablecaption{St Solutions for HL Tau Ring Masses from \citet{Pinte2016} vs \citet{Liu2017} \label{tab:HLTau compare}}
\tabletypesize{\scriptsize}
\tablewidth{0pt}
\tablehead{
\colhead{$\epsilon_{\rm trap}$} & \colhead{Rings} & \colhead{$f(>M_{\rm dust})_Q$} & \colhead{$\rm St_Q$} & \colhead{$f(>M_{\rm dust})_M$} & \colhead{$\rm St_M$} & \colhead{$f(>M_{\rm dust})_Q$} & \colhead{$\rm St_Q$} & \colhead{$f(>M_{\rm dust})_M$} & \colhead{$\rm St_M$} &  \\
\colhead{} & \colhead{} & \colhead{(Pinte)} & \colhead{(Pinte)} & \colhead{(Pinte)} & \colhead{(Pinte)} & \colhead{(Liu)} & \colhead{(Liu)} & \colhead{(Liu)} & \colhead{(Liu)} \\ 
\colhead{(1)} & \colhead{(2)} & \colhead{(3)} & \colhead{(4)} & \colhead{(5)} & \colhead{(6)} & \colhead{(7)} & \colhead{(8)} & \colhead{(9)} & \colhead{(10)}
}
\startdata
0.3 & B-49 & 72 & $7.86 \times 10^{-4}$ & 42 & $2.51 \times 10^{-3}$ & 89 & $3.45 \times10^{-4}$ & 27 & $3.30 \times 10^{-3}$ \\
 & B-72 & 72 & $2.45 \times 10^{-4}$ & 42 & $8.83 \times 10^{-4}$ & 89 & $2.42 \times 10^{-4}$ & 27 & $2.63 \times 10^{-3}$ \\
 & B-85 & 72 & $2.65 \times 10^{-3}$ & 42 & $7.60 \times 10^{-3}$ & 89 & $7.58 \times 10^{-4}$ & 27 & $7.31 \times 10^{-3}$ \\
0.6 & B-49 & 37 & $8.14 \times 10^{-4}$ & 18 & $2.81 \times 10^{-3}$ & 69 & $3.36 \times 10^{-4}$ & 3 & $4.20 \times 10^{-3}$ \\
 & B-72 & 37 & $2.59 \times 10^{-4}$ & 18 & $9.71 \times 10^{-4}$ & 69 & $2.44 \times 10^{-4}$ & 3 & $3.12 \times 10^{-3}$ \\
 & B-85 & 37 & $2.96 \times 10^{-3}$ & 18 & $8.11 \times 10^{-3}$ & 69 & $7.82 \times 10^{-4}$ & 3 & $7.08 \times 10^{-3}$ \\
1.0 & B-49 & 29 & $7.52 \times 10^{-4}$ & 3 & $3.39 \times 10^{-3}$ & 44 & $3.43 \times 10^{-4}$ & 1 & $5.86 \times 10^{-3}$ \\
 & B-72 & 29 & $2.49 \times 10^{-4}$ & 3 & $1.141 \times 10^{-3}$ & 44 & $2.50 \times 10^{-4}$ & 1 & $4.03 \times 10^{-3}$ \\
 & B-85 & 29 & $3.02 \times 10^{-3}$ & 3 & $6.70 \times 10^{-3}$ & 44 & $8.03 \times 10^{-4}$ & 1 & $8.19 \times 10^{-3}$ \\
\enddata
\tablecomments{Column(1): trapping efficiency $\epsilon_{\rm trap}$. Column(2): name of the ring. Column(3): maximum initial dust mass percentile determined by gravitational stability for \citet{Pinte2016}. Column(4): corresponding minimum St. Column(5): minimum initial dust mass percentile determined by the limit on the outer edge of the initial disk ($\rm a_0 < 1000 \rm AU$) for \citet{Pinte2016}. Column(6): corresponding maximum St. Columns (7)--(10): same as Columns (3)--(6) but for \citet{Liu2017}.}
\end{deluxetable*}

\subsection{Ring Masses} \label{subsec:Ring Mass}

We describe here how we compute the masses of the rings for which we apply our own Gaussian fitting.
To obtain the ring surface density $\Sigma$, we follow the radiative transfer equation:
\begin{equation}
    \text{I}_{\nu} = \Sigma\kappa\text{B}_{\nu},
    \label{eqn:Radiative Transfer}
\end{equation}
where $\text{I}_{\nu}$ is the specific intensity, $\Sigma$ is surface density, $\kappa$ is opacity, and $\text{B}_{\nu}$ is the Planck blackbody function evaluated at the disk midplane temperature. 
From DSHARP Mie-Opacity Library  \citep{Birnstiel2018}, we obtain $\kappa$ for grain sizes close to 1 mm and wavelength $\lambda$ at 0.87 mm, 0.9 mm, 1.25 mm, 1.3 mm, 2.1 mm, and 2.9 mm by modeling $\kappa ( \lambda )$ and interpolating with respect to $\lambda$ at which each system is measured at. We obtain the following $\kappa$ accordingly: $\kappa = $3.18 for 0.87 mm, 3.20 for 0.9 mm, 2.22 for 1.25 mm, 2.08 for 1.3 mm, 1.49 for 2.1 mm, and 0.75 for 2.9 mm. We obtain $\text{I}_\nu$ from our Gaussian radial intensity model parameters.

After obtaining the surface density for each ring, we integrate with respect to orbital distance over $\rm a_{\rm ring}\pm4\sigma$, assuming the ring to be optically thin:
\begin{equation}
    M_{\rm d}^{\rm thin} = \int_{\rm a_{\rm ring}-4\sigma}^{\rm a_{\rm ring}+4\sigma} 4\pi \rm r\Sigma \text{ dr}.
    \label{eqn:Surface density integral}
\end{equation}
To mitigate the effect of high optical depth (which can alter the shape of the apparent intensity profile compared to the true values), we calculate the true width $w_{\rm d}^{\rm true}$ using equation 35 from \citet{Dullemond2018}:
\begin{equation}
\begin{split}
    \frac{w_{\rm d}^{\rm mimick}}{w_{\rm d}^{\rm true}} \simeq W = \sqrt{2.15\ln \left( 1 + 0.148 \tau_{\nu}^{\rm peak}\right) + 1}, \\   
\end{split}
\end{equation}
where $w_{\rm d}^{\rm mimick}$ is the numerically obtained dust ring width through Gaussian fitting, $w_{\rm d}^{\rm true}$ is the true dust ring width, and $\tau_{\nu}^{\rm peak}=-\ln \left( 1 - \rm I_\nu/B_\nu \right)$ is the peak of the real optical depth profile obtained by setting $\rm I_\nu$ to the de-convolved Gaussian amplitude from our model fitting. We finally obtain the true dust ring mass accounting for the optical depth following equation 36 of \citet{Dullemond2018}:
\begin{equation}
\begin{split}
    M_{\rm d}^{\rm true} = M_{\rm d}^{\rm thin} \frac{1}{W} \frac{\tau_{\nu}^{\rm peak}}{1-e^{-\tau_{\nu}^{\rm peak}}}.  
\end{split}
\end{equation}

For AS 209, Elias 24, HD 163296, and GW Lup, we derive dust ring masses that are systematically lower than those reported by \citet{Dullemond2018}, by a factor of approximately 2. This discrepancy is primarily attributable to the larger flaring angle adopted in our analysis, 0.25 instead of 0.02, as well as to our use of a more precise opacity value, 2.22 instead of 2.00 $\mathrm{cm^2\,g^{-1}}$. The fitted parameters for these systems also differ modestly, reflecting differences in the adopted ring-fitting ranges and methodology. In addition, the beam dimensions quoted and used for Elias 24 in \citet{Dullemond2018}, $37\times 24$ mas, are inconsistent with those reported in \citet{Andrews2018b} and found in the FITS files for the observations, $37\times 34$ mas, the latter of which is what we adopt.

For the system HL Tau, we choose to take the ring masses from \citet{Pinte2016}, where dust emissions were found to be marginally optically thick at 2.9 mm. By contrast, \citet{Liu2017} found that dust emission for the rings B40 (B2 in \citet{ALMAPartnership2015}), B49 (B3 in \citet{ALMAPartnership2015}), and B85 (B6 in \citet{ALMAPartnership2015}) are optically thin at 2.9 and 7 mm due to their different choice of dust properties, effectively yielding lower ring masses for B-40 and B-85 and a higher ring mass for B-72. For our analysis, we choose to take values cited in \citet{Pinte2016} as our default but investigate how the alternative reporting of \citet{Liu2017} affect our solutions for St and $\alpha$.

\section{Results} \label{sec:Results}

\begin{figure*}
    \centering
    \begin{subfigure}[b]{18 cm}
        \centering
        \includegraphics[width=18 cm]{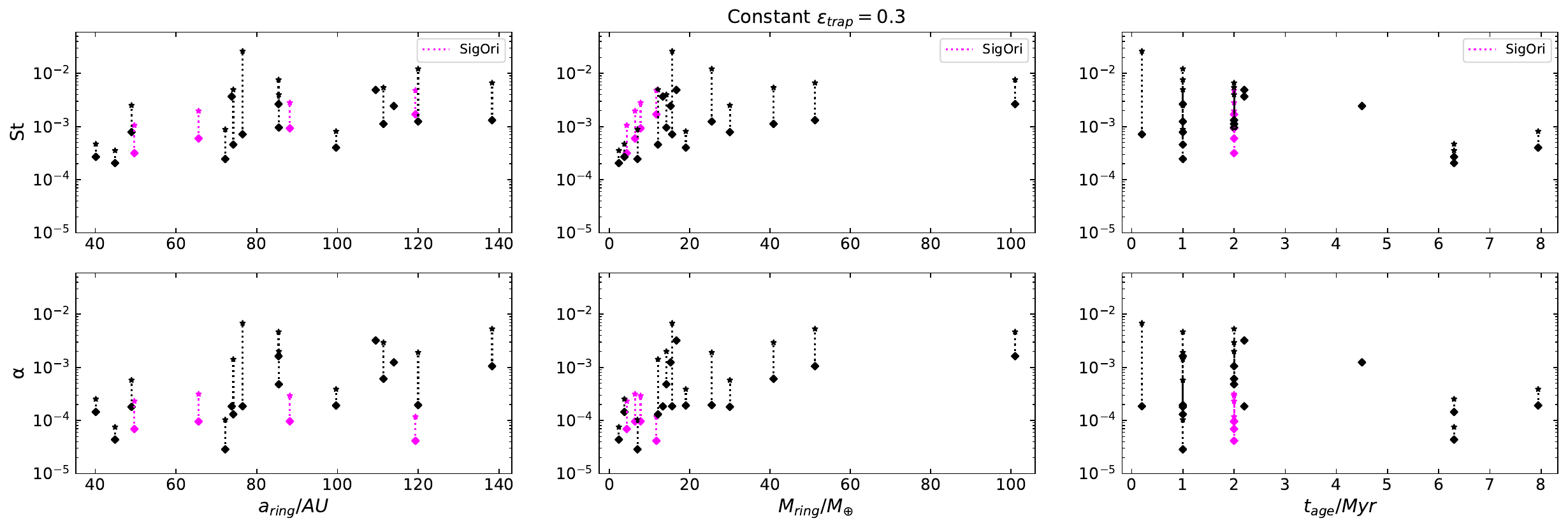}
    \end{subfigure}

    \begin{subfigure}[b]{18 cm}
        \centering
        \includegraphics[width=18 cm]{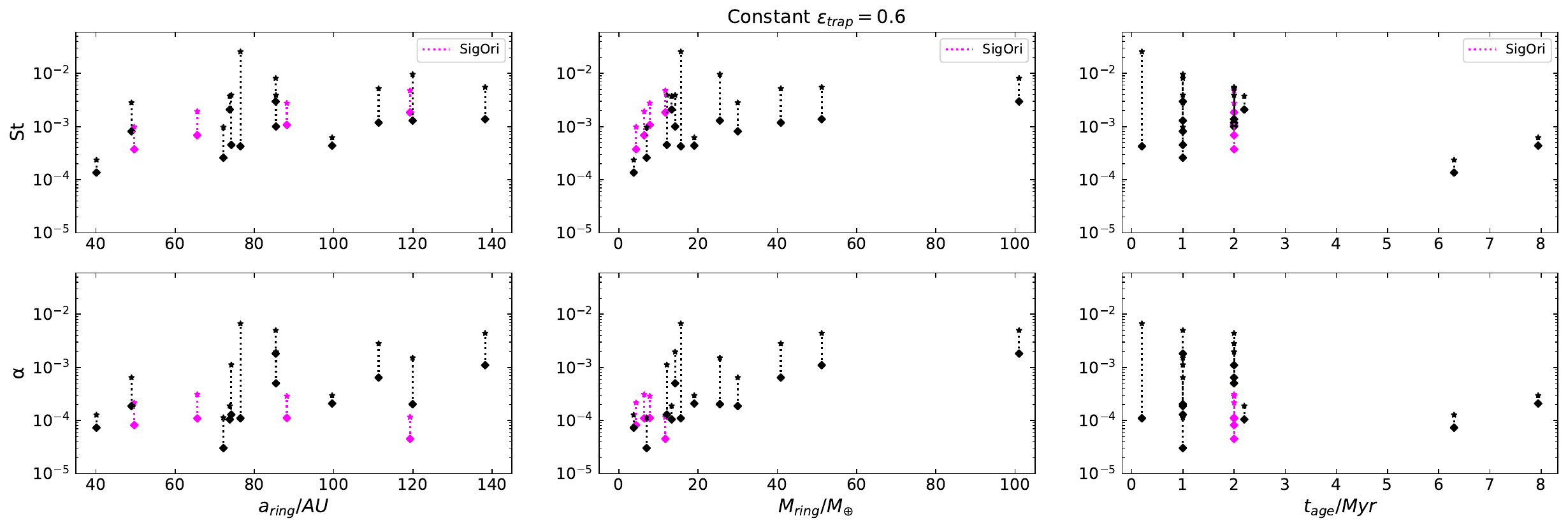}
    \end{subfigure}
    
    \begin{subfigure}[b]{18 cm}
        \centering
        \includegraphics[width=18 cm]{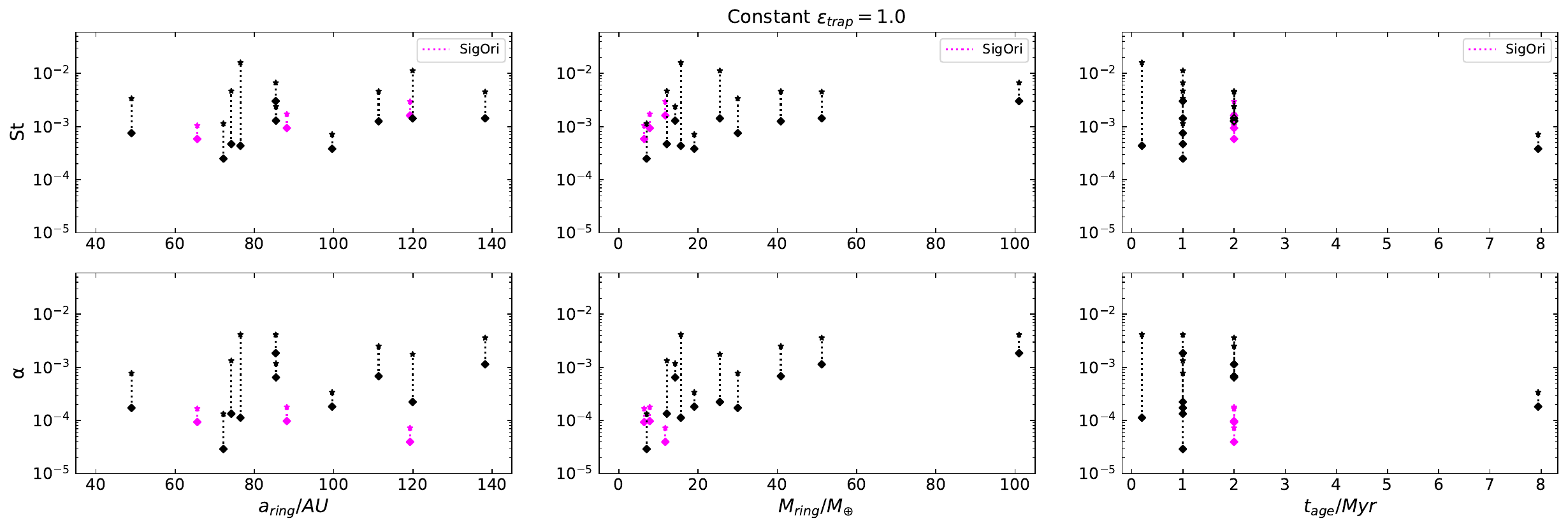}
    \end{subfigure}
    
    \caption{Stokes number St and turbulent $\alpha$ vs ring location (left), ring mass (middle), and system age (right) at $\epsilon = 0.3$ (top), 0.6 (middle), and 1.0 (bottom). The stars mark the upper limit of St \& $\alpha$ set by maximum $a_0=1000$ au, the diamonds mark the lower limit of St \& $\alpha$ set by the Toomre-Q stability, and magenta data points represent $\sigma$ Ori. We observe a generally low St and $\alpha$ values, ranging $10^{-4} < \rm St < 10^{-2}$ and $10^{-5} < \alpha < 10^{-2}$. Weak trends are observed: St and $\alpha$ increase with larger $a_{\rm ring}$ and larger $M_{\rm ring}$ and decrease with longer $t_{\rm age}$. Note that at $\epsilon_{\rm trap}=0.6$, GO Tau B-111, TW Hya B-45, and 2M0412 B-114 fail to produce solutions and at $\epsilon_{\rm trap}=1.0$, GO Tau B-74, GO Tau B-111, TW Hya B-40, TW Hya B-45, 2M0412 B-114, and SO 1274 B-50 fail to produce solutions.}
    \label{fig:St vs alpha graphs}
\end{figure*}

In Figure \ref{fig:St vs alpha graphs}, we present our calculated St and $\alpha$ vs orbital distance, system age, and mass of the 19 rings for $\epsilon_{\rm trap}=$0.3, 0.6, and 1.0. The sample presented here is slightly more than twice as large as the sample of rings studied in \citet{Lee2024}. Even with our bigger sample, we largely recover the result of \citet{Lee2024}, finding generally low St $\sim 10^{-4}$--$10^{-2}$ and correspondingly low $\alpha$ although some of the upper limits can be as large as $\sim$10$^{-2}$.

Higher $\epsilon_{\rm trap}$ leads to smaller St and therefore $\alpha$ but the effect is mostly minor, as long as the solution can be found. For $\epsilon_{\rm trap}=0.3$, we find solutions for all the rings in our sample. For $\epsilon_{\rm trap} = 0.6$, GO Tau B-111, TW Hya B-45, and 2M0412 B-114 fail to produce solutions, and for $\epsilon_{\rm trap} = 1.0$ (perfect trap), GO Tau B-74, GO Tau B-111, TW Hya B-40, TW Hya B-45, 2M0412 B-114, and SO 1274 B-50 are unable to produce solutions. The reason for this failure is the corresponding St falling below the limit set by gravitational stability. The host star luminosity of GO Tau, TW Hya, and 2M0412 is comparatively low which lowers $c_s$ and increases the minimum St corresponding to gravitational stability (equation \ref{eq:St_Q}). In addition, the rings TW Hya B-40 and B-45, and SO 1274 B-50 have low dust mass which leads to smaller St for a given initial dust mass $M_{\rm dust}$ to ensure slow radial drift.

We observe the following three trends of St with respect to system properties. First, St increases with larger $a_{\rm ring}$, which is a reflection of St $\propto a$ (with logarithmic corrections; see equation \ref{eq:St_r}) from our assumption of constant grain size.
Second, St increases with $M_{\rm ring}$ which is expected since creating a more massive ring in a given time requires larger rate of dust influx. Third, St decreases with increasing system age, which again is expected since accounting for a given ring of given mass over a longer timescale means slower inward dust drift.

As mentioned in Section \ref{subsec:Ring Mass}, we report on the effects comparing ring mass data taken from \citet{Pinte2016} and \citet{Liu2017} on the St and $\alpha$ of HL Tau. First, \citet{Liu2017} adopts 1.7$M_\odot$ as the mass of the host compared to the fiducial 1.4$M_\odot$ which lowers St$_Q$. Second, \citet{Liu2017} quote larger $M_{\rm ring}$ for B-72 ($\sim$17$M_\oplus$ vs.~7$M_\oplus$) and smaller $M_{\rm ring}$ for B-85 ($\sim$65$M_\oplus$ vs.~101$M_\oplus$). Larger ring mass leads to larger St in general as the dust grains would have to drift in faster to account for more massive rings whereas the opposite is true for smaller ring mass. For a given system, the two competing effects lead to only mild changes in the final maximum limit on the St. The ranges of St and initial disk mass percentile we obtain for two cases are listed in Table \ref{tab:HLTau compare}. 

Among the systems we study, GO Tau and 2M0412 have low mass host star with masses 0.36 $\rm M_{\odot}$ for GO Tau and 0.3 $\rm M_{\odot}$ for 2M0412. \citet{Andrews2018a} found a a weak positive correlation between the continuum effective size and the mass of the host star. Such correlation may be interpreted as low mass hosts preferentially hosting more compact disk, which is expected from the more rapid radial drift of grains around low mass stars \citep{Pinilla13}. However, GO Tau B-74, B-111 and 2M0412 B-114 are wide and their masses are run-of-the-mill compared to the other sharp rings in our sample. The only difference is that the hosts are low mass and less luminous, both of which are the key reason why dust grains are expected to drift in faster around low mass stars. The consequence is that for the fixed maximum outer edge of the initial disk, the maximum possible St will be lower around low mass stars. This expectation is borne out in our result that we are able to find self-consistent solutions for GO Tau and 2M0412 only when $\epsilon_{\rm trap}$ is small since for higher $\epsilon_{\rm trap}$, St$_r$ consistent with $a_0=1000$ au would imply Toomre-Q unstable disk. 

We do not see obvious peculiarity with the disks in $\sigma$ Ori in terms of our derived St and $\alpha$ other than what appears to be generally lower $\alpha$ than other systems for given $a_{\rm ring}$ (see the first column of Figure \ref{fig:St vs alpha graphs}). Given that their St are comparable to other systems all else equal, our result is a reflection of higher St/$\alpha$ of $\sigma$ Ori rings, suggesting these rings are particularly sharp and well-collected. We further note that the rings in $\sigma$ Ori have systematically lower $M_{\rm ring}$ than other systems. Even though the effect of external flux in the {\it bolometric} sense is minimal, the external UV flux is significantly elevated in $\sigma$ Ori compared to other systems in our sample \citep{Huang2024} which can lead to premature truncation of the gas disk by external photoevaporation \citep[e.g.,][]{Winter22,Hallatt25}. Given how the distribution of $a_{\rm ring}$'s are the same in $\sigma$ Ori as other systems, however, we do not see an evidence of systematic premature truncation of the gas disk. External photoevaporation may nevertheless cause gradual mass loss by winds that could carry away the grains; alternatively, the decrease in gas content can enhance local dust-to-gas ratio triggering clumping and dust coagulation into larger bodies, which will be invisible in mm wavelength.

\section{Discussion and Conclusion} \label{sec:disc_concl}

We collect a sample of 19 well-resolved, sharp rings in 10 systems and derive a range of St and $\alpha$ following the methodology of \citet{Lee2024} which reconciles the observed shape of the dust ring with the measurements of Class 0/I disk masses, representing the {\it initial} dust mass. Our sample is the largest to date for systematically deriving St and $\alpha$, breaking the degeneracy between the two parameters. We find the rings we study show generally low St ranging $10^{-4}$--$10^{-2}$ and equally low $\alpha$ that is $\sim$1--10 times smaller, largely reproducing earlier results that ringed disks are consistent with disks rich in gas (low St) and with low level of turbulence (low $\alpha$). We close with discussions of implications of our results.

\subsection{Stability Against Clumping} \label{subsec:clumping instability}

\begin{figure}[htbp]
    \centering
    \begin{subfigure}[b]{7 cm}
        \centering
        \includegraphics[width=7 cm]{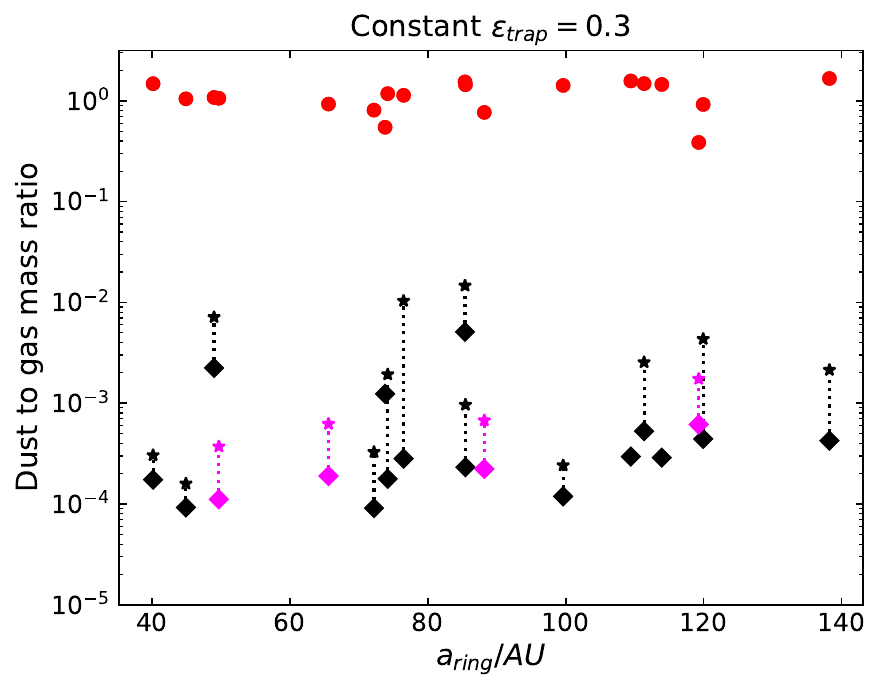}
    \end{subfigure}
    \begin{subfigure}[b]{7 cm}
        \centering
        \includegraphics[width=7 cm]{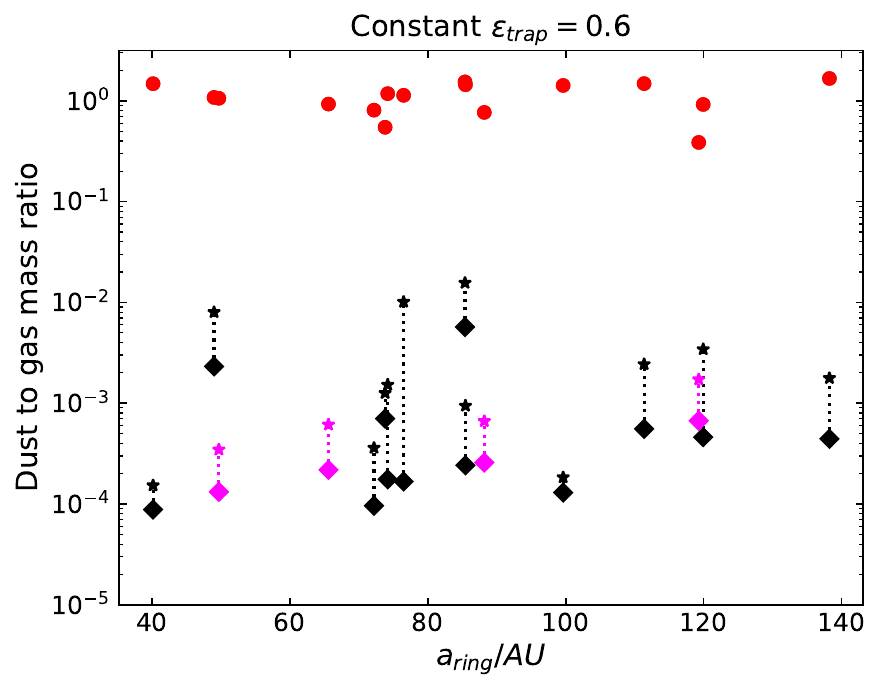}
    \end{subfigure}
    \begin{subfigure}[b]{7 cm}
        \centering
        \includegraphics[width=7 cm]{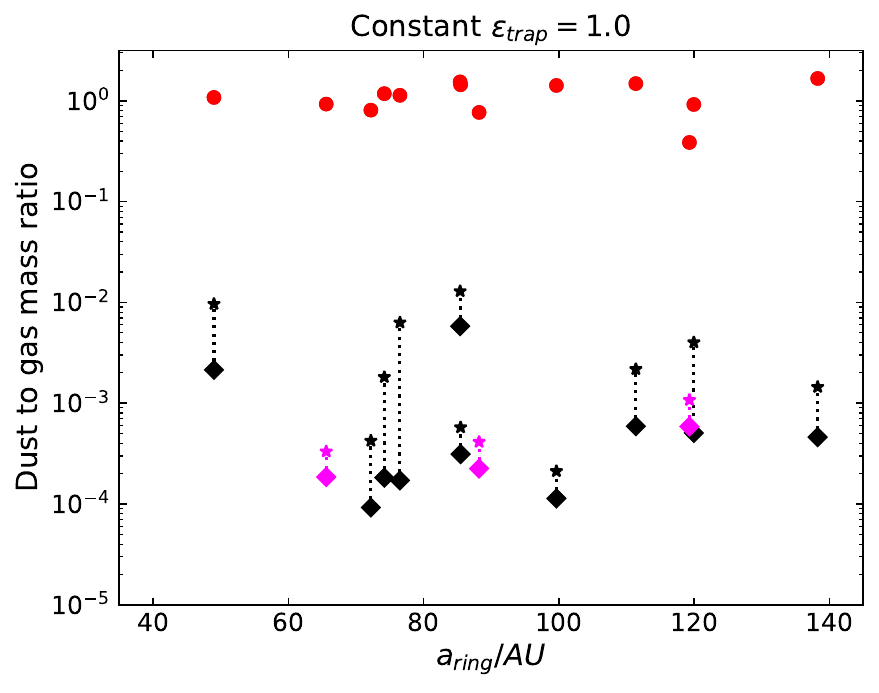}
    \end{subfigure}
    
    \caption{Local dust to gas mass ratio for minimum (diamonds) and maximum (stars) St. The red points are the critical dust to gas mass ratio to trigger streaming instability derived using Equation (14) from \citet{Li2021}, taking into account the effec of turbulence. All rings in this study are stable against clumping via streaming instability.}
    \label{fig:clumping stability}
\end{figure}

For a grain of given size, low St implies large $\Sigma_{\rm gas}$ and therefore likely a low dust-to-gas ratio. Such environments would be challenging for the clumping of solids via e.g., streaming instability (SI) in the non-linear regime. We verify whether the rings we study would be viable sites of clump formation. 

First, we calculate the ring solid surface density:
\begin{equation}
    \Sigma_{\rm ring} \equiv \frac{\rm M_{\rm ring}}{\left( 2 \pi \right)^{3/2}a_{\rm ring}w_d},
    \label{eq:Surface Density Ring}
\end{equation}
%
and the gas surface density using
\begin{equation}
    \Sigma_{\rm g} = \frac{\rho_s s}{\rm St}
\end{equation}
using the same $\rho_s$ and $s$ as equation \ref{eq:St_Q}.
We compute the maximum and minimum dust-to-gas mass ratio $\Sigma_{\rm ring}/\Sigma_{\rm g}$ corresponding to the maximum and minimum St, respectively, for each ring. 

Next, we consider the critical dust-to-gas mass ratio $Z_{\rm crit}$ that can trigger clumping by SI using equation 14 from \citet{Li2021}:
\begin{equation}
    Z > Z_{\rm crit} \equiv \epsilon_{\rm crit}(\rm St) \sqrt{H_{\rm p, \eta}^2 + H_{p, \alpha}^2}
\end{equation}
with $\epsilon_{\rm crit}(\rm St) = 2.5$ \citep[][equation 11]{Li2021} the critical midplane dust-gas ratio, $H_{\rm p, \eta} = h_\eta \Pi \simeq \Pi/5$ the scale height from pure dust-gas interaction with $\Pi = \left( v_{hw}/c_s\right)e^{-1/2}\gamma$ the global radial pressure gradient, $v_{hw} = 1/2(c_s^2/v_k)\gamma$ is the azimuthal headwind velocity, $h_\eta \simeq 0.2$ the acceleration due to our assumed disk gas profile, and $H_{\rm p, \alpha} = \sqrt{\alpha/\left( \alpha + \rm St \right)}$ the dust scale height considering additional turbulence. 
Consistent with discussion in \citet{Lee2024}, we find that the extra turbulence term dominates over the headwind term. Although our derived $\alpha$ is small, St/$\alpha$ is only a order of a few in most cases so the effect of turbulent stirring is enough to dilute the dust concentration in the midplane such that $Z$ of all of our rings fall significantly below $Z_{\rm crit} \sim 1$ for all values of $\epsilon_{\rm trap}$ we explore.

Our results imply that while the sharp, wide-orbit dust rings we study may be locally enhanced in dust compared to the background disk, they are not so enhanced that they would be efficient sites of clump formation. Such a finding is consistent with the fact that we observe these rings as rings, and it is self-consistent with our results of relatively low St which itself implies these disks remain gas-heavy. 

\subsection{Uncertainties with Disk and Ring Masses}

Our analysis depends on our adoption of Class 0/I disk masses from \citet{Tobin20}. We deliberately choose to use their Very Large Array 9 mm measurement where the emission is expected to be optically thin to ensure we capture the true initial disk dust mass. However, their derived dust mass remains dependent on their assumed disk thermal structure. Relaxing the isothermal assumption and allowing for the disk to be marginally gravitationally stable, \citet{Xu22}, using the same data given in \citet{Tobin20}, found the true Class 0/I disk masses may be elevated by roughly factors of 7. 

Increasing $M_0$ by a factor of 7 in equation \ref{eqn:Mass in Ring at time t} lowers our derived St for all the rings which is expected since to account for the same $M_{\rm ring}$ and at a given age, the radial drift would have to slow down. For many rings however, the resulting St is so low that the disk becomes Toomre-Q unstable---which is also expected given that the increase of $M_0$ simulates a disk that is only marginally stable. Specifically, we find that at $\epsilon_{\rm trap} = 0.3$ and 0.6, GOTau B-74, GOTau B-111, HD163296 B-100, TWHya B-40, TWHya B-45, GWLup B-86, 2M0412 B-114, and all of Sig Ori fail to produce solutions. For $\epsilon_{\rm trap} = 1.0$, all aforementioned rings, HLTau B-72, DoAr25 B-111, DoAr25 B-137, AS209 B-74, and AS209 B-120 fail to produce solutions. 

Another source of uncertainty is potential cluster-dependence on the initial mass budget. While we use the Class 0/I disk mass measurements from Orion, the ringed disks we study are found in a variety of environments. We consider this particular uncertainty to be mitigated by our use of fully scanning the entire range of Class 0/I disk masses which spans roughly $\sim$1.5 orders of magnitude, and our upper and lower limits are set by independent physical constraints: Toomre Q stability places an upper limit on $M_0$ and the physical extent of the disk places a lower limit. By contrast, the cluster-to-cluster variation in disk masses is observed to be within order unity---the median mass of Class 0 disks in Perseus is $\sim$158$M_\oplus$ \citep{Tychoniec20} compared to $\sim$200$M_\oplus$ in Orion \citep{Tobin20}.

Likewise, derivation of ring masses is subject to the assumed opacity and disk thermal structure. As described in Section \ref{subsec:Ring Mass}, our choice of larger flaring angle and more precise opacity give rise to $M_{\rm ring}$ that is systematically smaller (by factors of $\sim$2) than that listed in \citet{Dullemond18} which generally lowers St but within only a few \% given that $M_{\rm ring}$ enters in the logarithm in the calculation of St (equation \ref{eq:St_r}). Our method therefore is robust against order unity uncertainty in the ring masses.

\subsection{Case of Leaky Traps}

In general, we find that rings around low mass hosts or those with lower ring masses are better explained with leakier traps. It is not surprising that leaky traps would end up with less massive rings so we focus our discussion on the implication of disks around low mass hosts likely harboring rings of low trapping efficiency. Performing a simple radial dust evolution calculation in unperturbed and perturbed disk gas profile, \citet{Pinilla20} find that in order to explain the observed relationship between disk dust mass and host stellar mass, strong traps are necessary for stars more massive than the Sun. They find more ambiguous result for lower mass stars, concluding that any kind of traps (or no traps) can explain the observed trend. 

If the rings are generated by a planet, lower trapping efficiency around lower mass stars is qualitatively consistent with the current exoplanet observations. In particular, at $\gtrsim$1 au, giants---which would be able to carve a deep gap and create a strong trap---are known to be more abundant around high mass stars \citep[e.g.,][]{Nielsen19,Fulton21}. In addition, generally more massive planets are required to carve out a gap in puffier (i.e., larger aspect ratio) disks. At wide orbits where the heating is irradiation-dominated, the aspect ratio is $c_s/\Omega a \propto L_\star^{1/8}/M_\star^{1/2} \propto M_\star^{-5/16}$ for $L_\star \propto M_\star^{1.5}$ \citep{Choi16,Dotter16}. The empirical rarity of giants and the need for even more massive planets to carve a deep gap around low mass stars is consistent with leakier traps in a way that even if there was a planet generating a pressure bump, the bump will not be strong enough to completely or strongly trap the dust.
While tentatively supportive, our conclusion does not rule out other origin hypotheses for the dust traps as their trapping efficiency and incidence with respect to stellar mass remains largely unknown---more attention on non-planet hypotheses would be welcome.

In this study, we modeled the dust ring as a structure in steady state balance between aerodynamic drag and turbulent diffusion. The same steady state balance is widely used in the literature including seminal studies of dust filtration (synonymous with leakage) through a trap \citep[e.g.,][]{Zhu12}. Recent numerical studies suggest that this steady state balance may not be an accurate description of the dust-gas dynamics through planet-generated gas pressure bump \citep[e.g.,][]{Lee2022a,Huang2025}. Deriving an alternative semi-analytic and empirical description is a subject of ongoing work (R.~Li \& E.~J.~Lee, in preparation).

\begin{acknowledgments}
OW acknowledges support from the SPURS 2025 award and General Atomics. OW acknowledges the use of Claude AI as a coding assistant, especially for the iterative calculation of $M_{\rm 0,new}$. LC was supported by NSERC through the Undergraduate Summer Research Award. LC acknowledges the use of Composer 2.5 coding model in Cursor Pro code editor when investigating the cause of the discrepancy in DSHARP ring parameters. EJL was supported by NSF Research Grant 2509275, NSERC Discovery Grant RGPIN-2020-07045, DGECR-2020-00230, FRQNT/NSERC NOVA Grant FRQ-NT 2023-NOVA-325929, NSERC ALLRP 577027-22, and the William Dawson Scholarship from McGill University.
\end{acknowledgments}

\begin{contribution}
OW carried out the St and $\alpha$ calculations and wrote the initial draft. LC analyzed the radial intensity profiles, performed the initial calculations and contributed to the writing. EJL conceived the project, supervised the calculations, contributed to and reviewed the writing.
\end{contribution}

\appendix
\counterwithin{table}{section}
\counterwithin{figure}{section}

\section{Effect of Grain Growth}
\label{sec:app-grain-growth}

\begin{figure*}
    \centering
    \includegraphics[width=\textwidth]{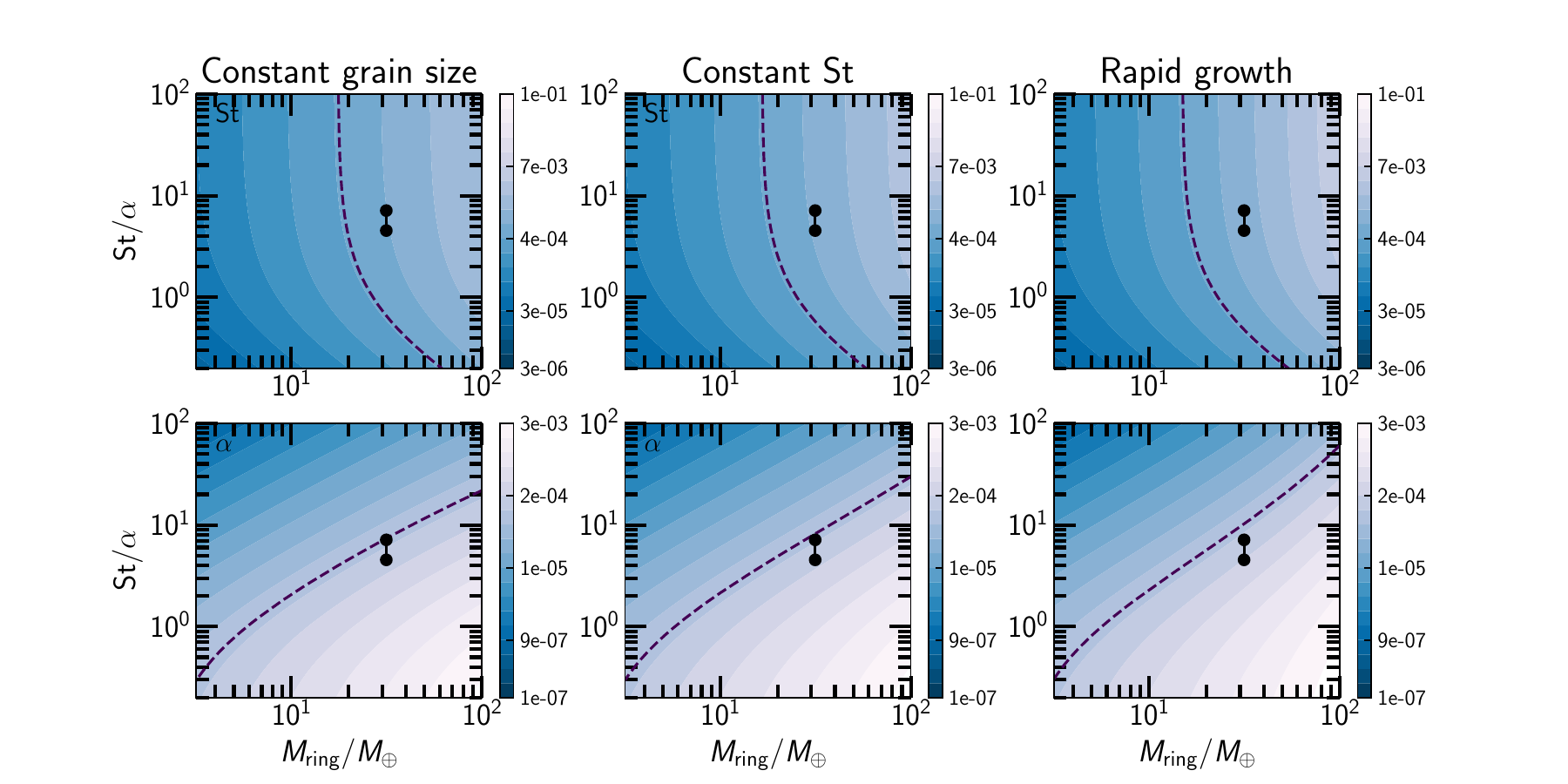}
    \caption{Contour map of St (top row) and $\alpha$ (bottom row) as a function of ${\rm St}/\alpha$ and the ring mass for a system parameter following AS209 B-74 with median $M_0$ and $\beta$. From left to right, we assume constant grain size, constant St, and St $\propto a^{-2}$, the last representing rapid grain growth. The purple dashed lines in the top row draw the minimum physical St that corresponds to Toomre $Q = 1$. The purple dashed line in the bottom row delineates $\alpha = 10^{-4}$ below which the disk is in the strongly inviscid limit. The measured St/$\alpha$ and $M_{\rm ring}$ for AS209 B-74 are indicated with black circles with the upper and lower limits connected by a line. We see negligible difference in the obtained St and $\alpha$ for all three cases.}
    \label{fig:grain-growth}
\end{figure*}

We have focused our analysis assuming constant grain size as the radio intensity profiles within a wavelength band more directly probe a given size of the grains. For grain fragmentation velocities $\gtrsim$10 cm/s and gas $\alpha \lesssim 10^{-3}$, the grains are expected to be drift-dominated with minimal size evolution at orbital distances beyond 10 au \citep{Chachan21} where the dust rings are. On the other hand, the presence of pressure trap modifies the dust motion and for efficient trapping will likely lead to fragmentation-growth balance.

\citet{Lee2024} showed that the final obtained St and $\alpha$ do not change whether spatially constant grain size or constant St is assumed. Constant St, in a sense, is consistent with the growth of grains as they drift in since St $\propto s/\Sigma_{\rm g} \propto sa$ such that $s \propto a^{-1}$. In Figure \ref{fig:grain-growth}, left and middle columns, we reproduce their result and verify that the result is robust to the two different assumptions. Depending on the microphysics of the dust grains, the grains may grow faster or slower (or fragment and become smaller). Since slower growth or more prominent fragmentation would simply lower St and $\alpha$ even more than what we find, we consider the case of faster grain growth.

To maintain the viability of analytic calculations, we choose St $\propto a^{-2}$. The dust equation of motion then reads:
\begin{align}
    \frac{\dot{a}}{a_{\rm ring}} &= -\frac{3}{2}\left(\frac{c_{s,r}}{v_{k,r}}\right)^2\frac{v_{k,r}}{a_{\rm ring}}\left[\alpha + \frac{2}{3}|\gamma|{\rm St}_r\left(\frac{a}{a_{\rm ring}}\right)^{-2}\right] \nonumber \\
    &\equiv -B\left[1 + \frac{2}{3}|\gamma|\frac{{\rm St}_r}{\alpha}\left(\frac{a}{a_{\rm ring}}\right)^{-2}\right]
\end{align}
Writing $x \equiv a/a_{\rm ring}$, the solution for $x$ is
\begin{equation}
    t'-t_{\rm age} = \frac{1}{B}\left[1-x(t') + C^{1/2}{\rm arctan}\left(\frac{C^{-1/2}(x(t')-1)}{1+x(t')/C}\right)\right],
    \label{eq:x}
\end{equation}
where $C = (2/3)|\gamma|{\rm St}_r/\alpha$ and $x(t_{\rm age})=1$. Combining with equation \ref{eqn:Mass in Ring at time t}, we have another boundary condition on $x$:
\begin{equation}
    x(0) = \left[\frac{M_{\rm ring}}{M_0\epsilon_{\rm trap}}\left(\frac{1\,{\rm au}}{a_{\rm ring}}\right)^\beta + 1\right]^{1/\beta}.
    \label{eq:x0}
\end{equation}
Plugging equation \ref{eq:x0} into equation \ref{eq:x}, we can isolate B:
\begin{equation}
    B = t_{\rm age}^{-1}\left[x(0)-1-C^{1/2}{\rm arctan}\left(\frac{C^{-1/2}(x(t')-1)}{1+x(t')/C}\right)\right],
\end{equation}
the solution of which will provide $\alpha$ and St$_r$ can be derived by multiplying the measurement-inferred St/$\alpha$ with the obtained $\alpha$.

The right column of Figure \ref{fig:grain-growth} presents the resulting St and $\alpha$ for a range of St/$\alpha$ and $M_{\rm ring}$ under St$\propto a^{-2}$. We see that the obtained St and $\alpha$ are remarkably consistent across all three assumptions from no growth to mild and to rapid growth of grains as they drift in. We therefore conclude that our method is robust to the detailed microphysics of dust grains, with a caveat that the a more detailed accounting of growth and fragmentation may deviate significantly from the three models we explored here, in which case the resultant $\alpha$ and St could be more different. Given our results however, we conjecture that it may be a second order effect.

\section{Limits on St and $\alpha$ for each ring}

We list here the adopted lower and upper limits of St and $\alpha$ for each of the sharp rings set by gravitational stability (subscript Q) and by the outer edge of the initial dust disk limited to 1000 au (subscript M), respectively. The corresponding Class 0/I disk mass percentiles are also listed. For systems with multiple rings, St and $\alpha$ are adjusted so that the same disk mass percentile is used (see Section \ref{subsec:St and Alpha} for more detail). Each table below shows results for $\epsilon_{\rm trap} = 0.3$ (Table \ref{tab:e_trap 0.3 list}), $\epsilon_{\rm trap} = 0.6$ (Table \ref{tab:e_trap 0.6 list}), and $\epsilon_{\rm trap} = 1.0$ (Table \ref{tab:e_trap 1.0 list}). Here, the zero mass percentile is defined as the minimum disk mass within each radial bin in the $M_{\rm disk}$ calculation while the 100th percentile is defined as the maximum disk mass within each radial bin.

\begin{deluxetable*}{lcCCCCCCC}
\tablewidth{0pt}
\tablecaption{Solutions for St and $\alpha$ with Limits from Section \ref{subsec:St and Alpha} at $\epsilon_{\rm trap} = 0.3$ \label{tab:e_trap 0.3 list}}
\tablehead{
\colhead{Source} & \colhead{Name} & \colhead{$f(>M_{\rm dust})_Q$} & \colhead{True $\rm St_Q$}  & \colhead{$\rm St_Q$} & \colhead{$\alpha_Q$} & \colhead{$f(>M_{\rm dust})_M$} & \colhead{$\rm St_M$} & \colhead{$\alpha_M$}\\
\colhead{} & \colhead{} & \colhead{(Percentile)} & \colhead{} & \colhead{} & \colhead{} & \colhead{(Percentile)} & \colhead{} & \colhead{}
}
\decimalcolnumbers
\startdata
GO Tau & B-74 & 1.0 & $2.00 \times 10^{-3}$ & $3.68 \times 10^{-3}$ & $1.84 \times 10^{-4}$ & 1.0 & $3.68 \times 10^{-3}$ & $1.84 \times 10^{-4}$ \\
GO Tau & B-111 & 1.0 & $3.98 \times 10^{-3}$ & $4.89 \times 10^{-3}$ & $3.21 \times 10^{-3}$ & 1.0 & $4.89 \times 10^{-3}$ & $3.21 \times 10^{-3}$ \\
HL Tau & B-49 & 58.0 & $1.23 \times 10^{-4}$ & $7.86 \times 10^{-4}$ & $1.80 \times 10^{-4}$ & 30.0 & $2.51 \times 10^{-3}$ & $5.75 \times 10^{-4}$ \\
HL Tau & B-72 & 58.0 & $2.42 \times 10^{-4}$ & $2.45  \times 10^{-4}$ & $2.90 \times 10^{-5}$ & 30.0 & $8.83 \times 10^{-4}$ & $1.03 \times 10^{-4}$ \\
HL Tau & B-85 & 58.0 & $3.24 \times 10^{-4}$ & $2.65 \times 10^{-3}$ & $1.62 \times 10^{-3}$ & 30.0 & $7.60 \times 10^{-3}$ & $4.66 \times 10^{-3}$ \\
DoAr 25 & B-111 & 51.0 & $9.01 \times 10^{-4}$ & $1.12 \times 10^{-3}$ & $6.08 \times 10^{-4}$ & 26.0 & $5.41 \times 10^{-3}$ & $2.93 \times 10^{-3}$ \\
DoAr 25 & B-137 & 51.0 & $1.32 \times 10^{-3}$ & $1.32 \times 10^{-3}$ & $1.05 \times 10^{-3}$ & 26.0 & $6.63 \times 10^{-3}$ & $5.29 \times 10^{-3}$ \\
Elias 24 & B-77 & 100.0 & $4.09 \times 10^{-4}$ & $7.15 \times 10^{-4}$ & $1.85 \times 10^{-4}$ & 1.0 & $2.62 \times 10^{-2}$ & $6.79 \times 10^{-3}$ \\
AS 209 & B-74 & 58.0 & $4.51 \times 10^{-4}$ & $4.57 \times 10^{-4}$ & $1.30 \times 10^{-4}$ & 2.0 & $4.96 \times 10^{-3}$ & $1.42 \times 10^{-3}$ \\
AS 209 & B-120 & 58.0 & $1.05 \times 10^{-3}$ & $1.24 \times 10^{-3}$ & $1.95 \times 10^{-4}$ & 2.0 & $1.22 \times 10^{-3}$ & $1.91 \times 10^{-3}$ \\
HD 163296 & B-100 & 26.0 & $3.56 \times 10^{-4}$ & $4.01 \times 10^{-4}$ & $1.91 \times 10^{-4}$ & 2.0 & $8.12 \times 10^{-4}$ & $3.87 \times 10^{-4}$ \\
TW Hya & B-40 & 1.0 & $1.30 \times 10^{-4}$ & $2.69 \times 10^{-4}$ & $1.45 \times 10^{-4}$ & 0.0 & $4.69 \times 10^{-4}$ & $2.52 \times 10^{-4}$ \\
TW Hya & B-45 & 1.0 & $1.58 \times 10^{-4}$ & $2.05 \times 10^{-4}$ & $4.40 \times 10^{-5}$ & 0.0 & $3.53 \times 10^{-4}$ & $7.50 \times 10^{-5}$ \\
GW Lup & B-86 & 29.0 & $9.30 \times 10^{-4}$ & $9.58 \times 10^{-4}$ & $4.79 \times 10^{-4}$ & 1.0 & $3.98 \times 10^{-3}$ & $1.99 \times 10^{-3}$ \\
2M0412 & B-114 & 1.0 & $2.23 \times 10^{-3}$ & $2.44 \times 10^{-3}$ & $1.24 \times 10^{-3}$ & 1.0 & $2.44 \times 10^{-3}$ & $1.24 \times 10^{-3}$ \\
SO 1274 & B-50 & 26.0 & $2.90 \times 10^{-4}$ & $3.16 \times 10^{-4}$ & $6.90 \times 10^{-5}$ & 1.0 & $1.05 \times 10^{-3}$ & $2.31 \times 10^{-4}$ \\
SO 1274 & B-66 & 26.0 & $4.70 \times 10^{-4}$  & $5.97 \times 10^{-4}$ & $9.50 \times 10^{-5}$ & 1.0 & $1.96 \times 10^{-3}$ & $3.14 \times 10^{-4}$ \\
SO 1274 & B-88 & 26.0 & $7.89 \times 10^{-4}$ & $9.27 \times 10^{-4}$ & $9.60 \times 10^{-5}$ & 1.0 & $2.810 \times 10^{-3}$ & $2.92 \times 10^{-4}$ \\
SO 1274 & B-119 & 26.0 & $1.33 \times 10^{-3}$ & $1.67 \times 10^{-3}$ & $4.10 \times 10^{-5}$ & 1.0 & $4.81 \times 10^{-3}$ & $1.17 \times 10^{-4}$ \\
\enddata
\tablecomments{Column(1): name of the source. Column(2): name of the ring. Column(3): maximum initial dust mass percentile determined by gravitational stability. Column(4): minimum St calculated from equation \ref{eq:St_Q} for each ring.
Column(5): adopted minimum St. Column(6): corresponding minimum $\alpha$. Column(6): minimum initial dust mass percentile determined by dust mass budget ($\rm a_0 < 1000 \rm AU$). Column(7): corresponding maximum St. Column(8): corresponding maximum $\alpha$.}
\end{deluxetable*}

\begin{deluxetable*}{lcCCCCCCC}
\tablewidth{0pt}
\tablecaption{Solutions for St and $\alpha$ with Limits from Section \ref{subsec:St and Alpha} at $\epsilon_{\rm trap} = 0.6$ \label{tab:e_trap 0.6 list}}
\tablehead{
\colhead{Source} & \colhead{Name} & \colhead{$f(>M_{\rm dust})_Q$} & \colhead{True $\rm St_Q$}  & \colhead{$\rm St_Q$} & \colhead{$\alpha_Q$} & \colhead{$f(>M_{\rm dust})_M$} & \colhead{$\rm St_M$} & \colhead{$\alpha_M$}\\
\colhead{} & \colhead{} & \colhead{(Percentile)} & \colhead{} & \colhead{} & \colhead{} & \colhead{(Percentile)} & \colhead{} & \colhead{}
}
\decimalcolnumbers
\startdata
GO Tau & B-74 & 1.0 &  $2.00 \times 10^{-3}$ & $2.09 \times 10^{-3}$ & $1.05 \times 10^{-4}$ & 0.0 & $3.74 \times 10^{-3}$ & $1.88 \times 10^{-4}$ \\
HL Tau & B-49 & 37.0 & $1.23 \times 10^{-4}$ & $8.14 \times 10^{-4}$ & $1.87 \times 10^{-4}$ & 18.0 & $2.81 \times 10^{-3}$ & $6.44 \times 10^{-4}$ \\
HL Tau & B-72 & 37.0 & $2.42 \times 10^{-4}$ & $2.59 \times 10^{-4}$ & $3.00 \times 10^{-5}$ & 18.0 & $9.71 \times 10^{-4}$ & $1.13 \times 10^{-4}$ \\
HL Tau & B-85 & 37.0 & $3.24 \times 10^{-4}$  & $2.96 \times 10^{-3}$ & $1.82 \times 10^{-3}$ & 18.0 & $8.11 \times 10^{-3}$ & $4.97 \times 10^{-3}$ \\
DoAr 25 & B-111 & 37.0 & $9.01 \times 10^{-4}$ & $1.18 \times 10^{-3}$ & $6.41 \times 10^{-4}$ & 3.0 & $5.18 \times 10^{-3}$ & $2.81 \times 10^{-3}$ \\
DoAr 25 & B-137 & 37.0 & $1.32 \times 10^{-3}$ & $1.38 \times 10^{-3}$ & $1.10 \times 10^{-3}$ & 3.0 & $5.50 \times 10^{-3}$ & $4.39 \times 10^{-3}$ \\
Elias 24 & B-77 & 89.0 & $4.09 \times 10^{-4}$ & $4.25 \times 10^{-4}$ & $1.10 \times 10^{-4}$ & 0.0 & $2.57 \times 10^{-2}$ & $6.65 \times 10^{-3}$ \\
AS 209 & B-74 & 38.0 & $4.51 \times 10^{-4}$ & $4.52 \times 10^{-4}$ & $1.29 \times 10^{-4}$ & 1.0 & $3.91 \times 10^{-3}$ & $1.12 \times 10^{-3}$ \\
AS 209 & B-120 & 38.0 & $1.05 \times 10^{-3}$ & $1.29 \times 10^{-3}$ & $2.02 \times 10^{-4}$ & 1.0 & $9.63 \times 10^{-3}$ & $1.51 \times 10^{-3}$ \\
HD 163296 & B-100 & 2.0 & $3.56 \times 10^{-4}$ & $4.37 \times 10^{-4}$ & $2.08 \times 10^{-4}$ & 1.0 & $6.19 \times 10^{-4}$ & $2.95 \times 10^{-4}$ \\
TW Hya & B-40 & 1.0 & $1.30 \times 10^{-4}$ & $1.36 \times 10^{-4}$ & $7.30 \times 10^{-5}$ & 0.0 & $2.35 \times 10^{-4}$ & $1.27 \times 10^{-4}$ \\
GW Lup & B-86 & 15.0 & $9.30 \times 10^{-4}$ & $1.00 \times 10^{-3}$ & $5.01 \times 10^{-4}$ & 0.0 & $3.89 \times 10^{-3}$ & $1.95 \times 10^{-3}$ \\
SO 1274 & B-50 & 2.0 & $2.90 \times 10^{-4}$ & $3.74 \times 10^{-4}$ & $8.20 \times 10^{-5}$ & 0.0 & $9.82 \times 10^{-4}$ & $2.15 \times 10^{-4}$ \\
SO 1274 & B-66 & 2.0 & $4.71 \times 10^{-4}$ & $6.86 \times 10^{-4}$ & $1.10 \times 10^{-4}$ & 0.0 & $1.93 \times 10^{-3}$ & $3.09 \times 10^{-4}$ \\
SO 1274 & B-88 & 2.0 & $7.89 \times 10^{-4}$ & $1.07 \times 10^{-3}$ & $1.11 \times 10^{-4}$ & 0.0 & $2.76 \times 10^{-3}$ & $2.87 \times 10^{-4}$ \\
SO 1274 & B-119 & 2.0 & $1.33 \times 10^{-3}$ & $1.85 \times 10^{-3}$ & $4.50 \times 10^{-5}$ & 0.0 & $4.74 \times 10^{-3}$ & $1.16 \times 10^{-4}$ \\
\enddata
\tablecomments{Same convention as Table \ref{tab:e_trap 0.3 list}.}
\end{deluxetable*}

\begin{deluxetable*}{lcCCCCCCC}
\tablewidth{0pt}
\tablecaption{Solutions for St and $\alpha$ with Limits from Section \ref{subsec:St and Alpha} at $\epsilon_{\rm trap} = 1.0$ \label{tab:e_trap 1.0 list}}
\tablehead{
\colhead{Source} & \colhead{Name} & \colhead{$f(>M_{\rm dust})_Q$} & \colhead{True $\rm St_Q$}  & \colhead{$\rm St_Q$} & \colhead{$\alpha_Q$} & \colhead{$f(>M_{\rm dust})_M$} & \colhead{$\rm St_M$} & \colhead{$\alpha_M$}\\
\colhead{} & \colhead{} & \colhead{(Percentile)} & \colhead{} & \colhead{} & \colhead{} & \colhead{(Percentile)} & \colhead{} & \colhead{}
}
\decimalcolnumbers
\startdata
HL Tau & B-49 & 29.0 & $1.23 \times 10^{-4}$ & $7.52 \times 10^{-4}$ & $1.72 \times 10^{-4}$ & 3.0 & $3.39 \times 10^{-3}$ & $7.77 \times 10^{-4}$ \\
HL Tau & B-72 & 29.0 & $2.42 \times 10^{-4}$ & $2.49 \times 10^{-4}$ & $2.90 \times 10^{-5}$ & 3.0 & $1.14 \times 10^{-3}$ & $1.33 \times 10^{-4}$ \\
HL Tau & B-85 & 29.0 & $3.24 \times 10^{-4}$ & $3.02 \times 10^{-3}$ & $1.85 \times 10^{-3}$ & 3.0 & $6.70 \times 10^{-3}$ & $4.10 \times 10^{-3}$ \\
DoAr 25 & B-111 & 29.0 & $9.01 \times 10^{-4}$ & $1.26 \times 10^{-3}$ & $6.81 \times 10^{-4}$ & 2.0 & $4.62 \times 10^{-3}$ & $2.51 \times 10^{-3}$ \\
DoAr 25 & B-137 & 29.0 & $1.32 \times 10^{-3}$ & $1.43 \times 10^{-3}$ & $1.14 \times 10^{-3}$ & 2.0 & $4.49 \times 10^{-3}$ & $3.59 \times 10^{-3}$ \\
Elias 24 & B-77 & 79.0 & $4.09 \times 10^{-4}$ & $4.35 \times 10^{-4}$ & $1.13 \times 10^{-4}$ & 0.0 & $1.60 \times 10^{-2}$ & $4.14 \times 10^{-3}$ \\
AS 209 & B-74 & 29.0 & $4.51 \times 10^{-4}$ & $4.69 \times 10^{-4}$ & $1.34 \times 10^{-4}$ & 0.0 & $4.67 \times 10^{-3}$ & $1.33 \times 10^{-3}$ \\
AS 209 & B-120 & 29.0 & $1.05 \times 10^{-3}$ & $1.42 \times 10^{-3}$ & $2.23 \times 10^{-4}$ & 0.0 & $1.13 \times 10^{-2}$ & $1.76 \times 10^{-3}$ \\
HD 163296 & B-100 & 1.0 & $3.56 \times 10^{-4}$ & $3.82 \times 10^{-4}$ & $1.82 \times 10^{-4}$ & 0.0 & $7.12 \times 10^{-4}$ & $3.39 \times 10^{-4}$ \\
GW Lup & B-86 & 1.0 & $9.30 \times 10^{-4}$ & $1.29 \times 10^{-3}$ & $6.46 \times 10^{-4}$ & 0.0 & $2.38 \times 10^{-3}$ & $1.19 \times 10^{-3}$ \\
SO 1274 & B-66 & 1.0 & $4.71 \times 10^{-4}$ & $5.83 \times 10^{-4}$ & $9.30 \times 10^{-5}$ & 0.0 & $1.04 \times 10^{-3}$ & $1.67 \times 10^{-4}$ \\
SO 1274 & B-88 & 1.0 & $7.89 \times 10^{-4}$ & $9.37 \times 10^{-4}$ & $9.70 \times 10^{-5}$ & 0.0 & $1.72 \times 10^{-3}$ & $1.79 \times 10^{-4}$ \\
SO 1274 & B-119 & 1.0 & $1.33 \times 10^{-3}$ & $1.62 \times 10^{-3}$ & $3.90 \times 10^{-5}$ & 0.0 & $2.96 \times 10^{-3}$ & $7.20 \times 10^{-5}$ \\
\enddata
\tablecomments{Same convention as Table \ref{tab:e_trap 0.3 list}.}
\end{deluxetable*}


\section{Complete Table of Rings in the Literature}

We list here in Tables \ref{tab:big data table} and \ref{tab:ODISEA Data} the full sample of systems with dust rings that we used before passing through our sharp ring criterion. They comprise of 48 systems in PP7, 4 systems in $\sigma$ Ori, 4 Taurus M dwarf systems (Table \ref{tab:big data table}). We separately list ODISEA rings in Table \ref{tab:ODISEA Data}.

\startlongtable
\begin{deluxetable*}{lcccccc}
\tablewidth{0pt}
\tabletypesize{\scriptsize}
\tablecaption{PP7, $\sigma$-Orionis, and Taurus M Stars Preprocessed Data \label{tab:big data table}}
\tablehead{
\colhead{Source} & \colhead{Name} & \colhead{$M_{\star}$} & \colhead{$L_{\star}$} & \colhead{$a_{\rm ring}$} & \colhead{$\sigma$} & \colhead{Reference} \\
\colhead{} & \colhead{} & \colhead{$(M_{\odot})$} & \colhead{$(L_{\odot})$} & \colhead{(au)} & \colhead{(au)} & \colhead{} 
}
\startdata
AATau & B-49 & 0.85 & 0.754 & 48.7 & 9.60 & 1,52 \\
AATau & B-95 & 0.85 & 0.754 & 94.9 & 12.2 & 1,52 \\
AATau & B-143 & 0.85 & 0.754 & 143.0 & 11.2 & 1,52 \\
CITau & B-28 & 0.89 & 0.81 & 27.7 & 19.3 & 2,3,4 \\
CITau & B-62 & 0.89 & 0.81 & 62.0 & 29.4 & 2,3,4 \\
CITau & B-99 & 0.89 & 0.81 & 99.2 & 8.7 & 2,3,4 \\
CITau & B-153 & 0.89 & 0.81 & 153.0 & 59.7 & 2,3,4 \\
CIDA1 & B-16 & 0.2 & 0.15 & 16.2 & 3.5 & 5,6 \\
CIDA1 & B-21 & 0.2 & 0.15 & 21.0 & 9.7 & 5,6 \\
CIDA9A & B-40 & 0.43 & 0.2 & 39.5 & 25.3 & 3,4,25,26 \\
DLTau & B-46 & 0.98 & 0.65 & 46.4 & 14.6 & 3,4 \\
DLTau & B-78 & 0.98 & 0.65 & 78.1 & 8.6 & 3,4 \\
DLTau & B-112 & 0.98 & 0.65 & 112.0 & 29.6 & 3,4 \\
DMTau & B-24 & 0.53 & 0.3 & 24.0 & 0.722 & 7,8,9 \\
DNTau & B-15 & 0.52 & 0.7 & 15.4 & 21.1 & 3,10 \\
DNTau & B-53 & 0.52 & 0.7 & 53.4 & 7.68 & 3,10 \\
DSTau & B-57 & 0.58 & 0.25 & 56.8 & 7.30 & 4,25 \\
FTTau & B-32 & 0.34 & 0.15 & 32.1 & 16.5 & 3,4 \\
GMAur & B-37 & 1.3 & 1.58 & 37.0 & 4.08 & 9,11,12 \\
GMAur & B-83 & 1.3 & 1.58 & 82.7 & 4.42 & 9,11,12 \\
GMAur & B-177 & 1.3 & 1.58 & 177.0 & 21.5 & 9,11,12 \\
GOTau & B-73 & 0.36 & 0.21 & 73.0 & 1.97 & 3,10 \\
GOTau & B-109 & 0.36 & 0.21 & 109.0 & 9.37 & 3,10 \\
IPTau & B-27 & 0.52 & 0.34 & 27.1 & 10.4 & 3,4,10 \\
IQTau & B-48 & 0.5 & 0.22 & 48.2 & 11.8 & 3,10 \\
IQTau & B-83 & 0.5 & 0.22 & 82.8 & 24.5 & 3,10 \\
IRAS04125+2902 & B-55 & 0.5 & 0.4 & 55.0 & 14.9 & 13 \\
LkCa15 & B-47 & 1.01 & 1.2 & 47.3 & 9.33 & 14,15,16 \\
LkCa15 & B-69 & 1.01 & 1.2 & 69.0 & 6.32 & 14,15,16 \\
LkCa15 & B-100 & 1.01 & 1.2 & 100.0 & 14.5 & 14,15,16 \\
MHO6 & B-65 & 0.17 & 0.06 & 64.7 & 18.3 & 6 \\
MWC480/HD31648 & B-98 & 1.91 & 17.8 & 97.6 & 12.56 & 3 \\
RYTau & B-18 & 2.04 & 12.3 & 18.2 & 25.6 & 3,4,17 \\
RYTau & B-49 & 2.04 & 12.3 & 49.0 & 19.5 & 3,4,17 \\
UXTauA & B-38 & 1.4 & 2.5 & 37.5 & 4.48 & 26,27,28,29 \\
HLTau & B-21 & 1.7 & 11.0 & 21.4 & 6.45 & 18,19 \\
HLTau & B-40 & 1.7 & 11.0 & 40.0 & 3.2 & 18,19 \\
HLTau & B-49 & 1.7 & 11.0 & 49.0 & 2.1 & 18,19 \\
HLTau & B-58 & 1.7 & 11.0 & 58.0 & 7.5 & 18,19 \\
HLTau & B-72 & 1.7 & 11.0 & 72.2 & 2.05 & 18,19 \\
HLTau & B-85 & 1.7 & 11.0 & 85.4 & 6.0 & 18,19 \\
CQTau & B-53 & 1.67 & 10.0 & 53.0 & 13.0 & 20,21,22 \\
V1247Ori & B-75 & 1.86 & 15.8 & 75.0 & 17.0 & 23,24,53 \\
SR4 & B-18 & 0.68 & 1.17 & 18.0 & 6.65 & 19,32 \\
GSS26/ISO-Oph17 & B-25 & 0.5 & 0.2 & 25.0 & 2.97 & 31,34 \\
GSS26/ISO-Oph17 & B-47 & 0.5 & 0.2 & 47.0 & 8.49 & 31,34 \\
Elias2-20 & B-29 & 0.48 & 2.24 & 29.0 & 2.6 & 19,32 \\
Elias2-20 & B-36 & 0.48 & 2.24 & 36.0 & 0.8 & 19,32 \\
DoAr25 & B-111 & 0.95 & 0.95 & 111.0 & 7.15 & 19,32 \\
DoAr25 & B-137 & 0.95 & 0.95 & 137.0 & 6.4 & 19,32 \\
Elias2-24 & B-77 & 0.78 & 6.02 & 76.7 & 6.1 & 19,32 \\
GSS39/Elias2-27 & B-86 & 0.49 & 0.91 & 86.0 & 10.6 & 19,30,31 \\
ISO-Oph196/WSB60 & B-8 & 0.2 & 0.2 & 8.0 & 3.40 & 32 \\
ISO-Oph196/WSB60 & B-34 & 0.2 & 0.2 & 34.0 & 6.79 & 32 \\
DoAr44 & B-47 & 1.4 & 1.8 & 47.0 & 5.52 & 30,32 \\
RXJ1633.9-2442 & B-36 & 0.8 & 1.0 & 36.0 & 7.64 & 30,32 \\
WSB82 & B-50 & 1.5 & 5.1 & 50.0 & 11.0 & 30,32 \\
WSB82 & B-123 & 1.5 & 5.1 & 123.0 & 10.6 & 30,32 \\
WSB82 & B-269 & 1.5 & 5.1 & 269.0 & 33.1 & 30,32 \\
AS209 & B-14 & 0.83 & 1.41 & 14.2 & 3.78 & 33 \\
AS209 & B-28 & 0.83 & 1.41 & 27.8 & 2.00 & 33 \\
AS209 & B-39 & 0.83 & 1.41 & 38.7 & 1.44 & 33 \\
AS209 & B-74 & 0.83 & 1.41 & 74.2 & 3.95 & 33 \\
AS209 & B-97 & 0.83 & 1.41 & 96.7 & 3.44 & 33 \\
AS209 & B-120 & 0.83 & 1.41 & 120.0 & 4.76 & 33 \\
AS209 & B-141 & 0.83 & 1.41 & 141.0 & 1.19 & 33 \\
SR24S & B-30 & 1.4 & 2.0 & 30.0 & 12.7 & 30,32,34 \\
SR21 & B-36 & 2.6 & 12.9 & 36.0 & 6.33 & 10, 35,36 \\
HD163296 & B-16 & 2.0 & 17.0 & 15.5 & 8.7 & 37 \\
HD163296 & B-67 & 2.0 & 17.0 & 67.1 & 6.56 & 37 \\
HD163296 & B-101 & 2.0 & 17.0 & 101.0 & 5.8 & 37 \\
HPCha & B-45 & 0.95 & 2.4 & 44.8 & 6.4 & 29,39 \\
WWCha & B-68 & 1.0 & 11.0 & 68.3 & 11.0 & 10,38 \\
WWCha & B-147 & 1.0 & 11.0 & 147.0 & 43.0 & 10,38 \\
TWHya & B-3 & 0.81 & 0.34 & 3.0 & 0.934 & 10,41,42 \\
TWHya & B-30 & 0.81 & 0.34 & 29.5 & 2.55 & 10,41,42 \\
TWHya & B-33 & 0.81 & 0.34 & 33.0 & 4.25 & 10,41,42 \\
TWHya & B-45 & 0.81 & 0.34 & 44.7 & 1.19 & 10,41,42 \\
PDS70 & B-74 & 0.76 & 0.36 & 73.7 & 14.8 & 43,44,52 \\
Sz82,IMLup & B-134 & 0.89 & 2.6 & 134.0 & 9.2 & 19,45 \\
HD142527 & B-205 & 0.63 & 1.44 & 205.0 & 29.7 & 46 \\
Sz83,RULup & B-24 & 0.54 & 0.19 & 24.0 & 4.0 & 19 \\
Sz83,RULup & B-34 & 0.54 & 0.19 & 34.0 & 2.75 & 19 \\
Sz91 & B-110 & 0.55 & 0.13 & 110.5 & 22.2 & 10, 45 \\
Sz123A & B-32 & 1.45 & 7.2 & 32.4 & 6.49 & 51 \\
SAO206462/HD135344B & B-51 & 1.78 & 3.8 & 51.3 & 8.1 & 10,47 \\
J1610 & B-29 & 0.67 & 0.46 & 28.8 & 2.42 & 16 \\
J1610 & B-41 & 0.67 & 0.46 & 41.1 & 3.36 & 16 \\
HD100546 & B-28 & 2.4 & 25.1 & 28.4 & 8.47 & 40 \\
HD100546 & B-199 & 2.4 & 25.1 & 199.0 & 23.3 & 40 \\
SO 844 & B-20 & 0.44 & 0.62 & 20.5 & 2.8 & 50 \\
SO 844 & B-37 & 0.44 & 0.62 & 36.8 & 3.0 & 50 \\
SO 897 & B-12 & 0.7 & 0.85 & 11.9 & 3.9 & 50 \\
SO 1153 & B-18 & 0.9 & 0.68 & 17.5 & 5.2 & 50 \\
SO 1274 & B-33 & 0.64 & 0.68 & 32.8 & 2.4 & 50 \\
SO 1274 & B-50 & 0.64 & 0.68 & 49.7 & 2.1 & 50 \\
SO 1274 & B-66 & 0.64 & 0.68 & 65.6 & 2.6 & 50 \\
SO 1274 & B-88 & 0.64 & 0.68 & 88.2 & 3.1 & 50 \\
SO 1274 & B-119 & 0.64 & 0.68 & 119.0 & 2.3 & 50 \\
2M0412 & B-62 & 0.3 & 0.126 & 61.7 & 17.5 & 48,49 \\
2M0412 & B-114 & 0.3 & 0.126 & 114.0 & 9.48 & 48,49 \\
2M0434 & B-10 & 0.3 & 0.115 & 9.82 & 8.19 & 48,49 \\
2M0436 & B-12 & 0.58 & 0.146 & 12.2 & 0.99 & 48,49 \\
2M0508 & B-7 & 0.44 & 0.11 & 6.79 & 7.15 & 48,49 \\
\enddata
\tablecomments{Literature search rings from PP7 \citep{Bae2023}, $\sigma$-Orionis \citep{Huang2024}, and Taurus M stars \citep{Shi2024, Long2023}. References: (1) \citet{Loomis2017}, (2) \citet{Clarke2018}, (3) \citet{Long2018}, (4) \citet{Long2019}, (5) \citet{Pinilla2021}, (6) \citet{Kurtovic2021}, (7) \citet{Hashimoto2021}, (8) \citet{Andrews2011}, (9) \citet{Isella2009}, (10) \citet{Andrews2018a}, (11) \citet{Hughes2009}, (12) \citet{Huang2020}, (13) \citet{Espaillat2015}, (14) \citet{Andrews2011}, (15) \citet{Isella2012}, (16) \citet{Facchini2020}, (17) \citet{Isella2010}, (18) \citet{ALMAPartnership2015}, (19) \citet{Huang2018}, (20) \citet{UbeiraGabellini2019}, (21) \citet{Uyama2020}, (22) \citet{Wolfer2021}, (23) \citet{Ohta2016}, (24) \citet{Kraus2017}, (25) \citet{Akeson2014}, (26) \citet{Pinilla2018}, (27) \citet{Menard2020}, (28) \citet{Zapata2020}, (29) \citet{Francis2020}, (30) \citet{Cieza2019}, (31) \citet{Simon2017}, (32) \citet{Cieza2021}, (33) \citet{Andrews2018b}, (34) \citet{Pinilla2017}, (35) \citet{Perez2014}, (36) \citet{Pinilla2015}, (37) \citet{Isella2018}, (38) \citet{Pascucci2016}, (39) \citet{Norfolk2021}, (40) \citet{Perez2020}, (41) \citet{Andrews2016}, (42) \citet{Tsukagoshi2016}, (43) \citet{Keppler2019}, (44) \citet{Isella2019}, (52) \citet{Benisty2021}, (45) \citet{Ansdell2018}, (46) \citet{Fukagawa2013}, (47) \citet{vanderMarel2016}, (48) \citet{Long2023}, (49) \citet{Shi2024}, (50) \citet{Huang2024}, (51) \citet{Guerra-Alvarado2025}, (52) \citet{Gudel2007}, (53) \citet{Ren2024}}
\end{deluxetable*}

\begin{deluxetable*}{lcccccccccccccccccc}[!h]
\tablewidth{0pt}
\tablecaption{ODISEA Rings \label{tab:ODISEA Data}}
\tablehead{
\colhead{Source} & \colhead{Name} & \colhead{$a_{\rm ring}$} & \colhead{Max $a_{\rm ring}$} & \colhead{Min $a_{\rm ring}$} & \colhead{$\sigma$} & \colhead{Max $\sigma$} & \colhead{Min $\sigma$} & \colhead{Reference} & \colhead{$L_\star$} & \colhead{$M_\star$} \\
\colhead{} & \colhead{} & \colhead{(au)} & \colhead{(au)} & \colhead{(au)} & \colhead{(au)} & \colhead{(au)} & \colhead{(au)} & \colhead{} & \colhead{($L_{\odot}$)} & \colhead{($M_{\odot}$)} \\
}
\startdata
ISO-Oph 17 & B-25 & 25 & 30 & 20 & 7 & 9 & 5 & 1 & 0.2 & 0.5 \\
ISO-Oph 17 & B-47 & 47 & 56 & 38 & 20 & 24 & 16 & 1 & 0.2 & 0.5 \\
DoAr 44 & B-47 & 47 & 49 & 45 & 13 & 15 & 11 & 2 & 1.8 & 1.4 \\
WSB 82 & B-50 & 50 & 52 & 48 & 26 & 28 & 24 & 3 & 5.1 & 1.5 \\
WSB 82 & B-123 & 123 & 126 & 120 & 25 & 27 & 23 & 3 & 5.1 & 1.5 \\
WSB 82 & B-269 & 269 & 274 & 264 & 78 & 80 & 76 & 3 & 5.1 & 1.5 \\
ISO-Oph 2A & B-49 & 49 & 53 & 45 & 11 & 13 & 9 & 4 & 0.7 & 0.5 \\
ISO-Oph 2A & B-69 & 69 & 74 & 64 & 13 & 15 & 11 & 4 & 0.7 & 0.5 \\
ISO-Oph 196 & B-8 & 8 & 10 & 6 & 8 & 10 & 6 & 2 & 0.2 & 0.2 \\
ISO-Oph 196 & B-34 & 34 & 36 & 32 & 16 & 18 & 14 & 2 & 0.2 & 0.2 \\
EM* SR 24S & B-30 & 30 & 32 & 28 & 30 & 32 & 28 & 5 & 2.0 & 1.4 \\
RXJ1633.9-2442 & B-36 & 36 & 37 & 35 & 18 & 19 & 17 & 6 & 1.0 & 0.8 \\
\enddata
\tablecomments{For ODISEA, we also check for both optimistic and pessimistic scenarios. The optimistic scenario is defined as when the ring location from the host star is the largest (Max $a_{\rm ring}$) and the width is the smallest (Min $\sigma$), while the pessimistic scenario is when the ring location is the smallest (Min $a_{\rm ring}$) and the width is the largest (Max $\sigma$). Min and Max $a_{\rm ring}$ and $\sigma$ are both obtained via the uncertainties for $a_{\rm ring}$ and $\sigma$ listed in \citet{Cieza2021} Table 6. In both the optimistic and pessimistic scenarios, no rings pass our dust ring width search criterion, $w_{\rm d} < \rm H/\sqrt{2}$, outlined in section \ref{subsec:Sample Selection}. References: (1) \citet{Ricci2010}, (2) \citet{Manara2014}, (3) Ruiz-Rodriguez (2025, in preparation), (4) \citet{Gatti2006}, (5) \citet{Natta2006}, (6) \citet{Cieza2012}.}
\end{deluxetable*}

\clearpage
\bibliography{dustring}{}
\bibliographystyle{aasjournalv7}

\end{document}